\documentstyle [12pt,epsf]{article}
\begin{document}
\newcommand{\vm}{\vspace{0.2cm}}
\newcommand{\vl}{\vspace{0.4cm}}

\title{About the construction of p-adic QFT limit of TGD}

\author{Matti Pitk\"anen
\\
\small \em Torkkelinkatu 21 B39 , 00530, Helsinki, Finland\\
}

\date{16. October 1995}

\maketitle

\newpage

\begin{center}
Abstract
\end{center}

\vm

The p-adic description of Higgs mechanism  provides excellent predictions
for elementary particle and hadron masses.  In this work the construction of
p-adic field theory limit of Quantum TGD is considered. Topological condensate
defined as a manifold obtained by gluing together p-adic spacetime regions
with different values of $p$ is identified as 'quantum average'   space time
defined as absolute minimum of effective action $S_e$ associated with K\"ahler
Dirac action defining configuration space geometry.  p-Adic spacetime
topology is ultrametric in accordance with the analogy with spin glass phase
implied by the vacuum degenerary of  K\"ahler action. Symmetry arguments
suggest that $S_e$ in presence of fermion fields is just the   super
symmetrized K\"ahler Dirac action.  Super symmetrization  motivated by  $N=1$
super symmetry generated by right handed neutrino replaces H-coordinates with
their super  counterparts and requires color singletness of physical states
appears as consistency condition. p-Adic $Diff(M^4)$ and canonical
transformations  of $CP_2$ localized with respect to $M^4$  are approximate
symmetries of the effective action and broken only by the nonflatness of the
induced metric. Super gauge invariances  in  interior allow only massless many
boson states with vanishing electroweak and color quantum numbers including
graviton in accordance with the hypothesis that elementary particle quantum
numbers reside on boundaries. Boundary components are idealized with world
lines and  Kac Moody Dirac spinors  used in mass calculations  describe the
conformal degrees of freedom of the  boundary component. Perturbation theory
in powers of $p$ converges extremely rapidly: UV divergences are absent but
natural infrared cutoff is present.

\newpage

\section{Introduction}

 The p-adic description of Higgs
mechanism  provides excellent predictions for
elementary particle and hadron masses  (hep-th@xxx.lanl.gov 9410058-62).
The decription was based on rather general assumptions motivated by Quantum
TGD. Canonical correspondence between p-adic and real numbers is an essential
element of the model. p-Adic thermodynamics gives mass squared as a thermal
expectation value and massivation results from the thermal excitation of
Planck mass states. p-Adic length scale hypothesis states
that fundamental basic length scales are given by $Lp=\sqrt{p}L_0$, $L_0\simeq
1.824 \cdot 10^4\sqrt{G}$ and that physically
most interesting primes correspond to primes near prime powers of two.
p-Adic super  conformal invariance at boundary components together with the
concept of Kac-Moody spinor are also cornerstone of the approach.
  The
 four-dimensionality of the
algebraic extension allowing the existence of p-adic square root in the
 vicinity of
p-adic real axis is essential element and
this  suggests that 4-dimensional p-adic conformal invariance might have key
role in the   the construction  of p-adic field
 theory
limit of TGD  but it turns out that the actual
approximate gauge symmetry in  the interior of spacetime contains
$Diff^4(M^4)$ and
 is therefore  much
larger than the group of p-adic conformal transformations ($Conf(M^4)$).
 What remained lacking from the spectrum was graviton. This
 is
understandable since only the boundary degrees of freedom of 3-surface were
 taken into
account in the calculation and it is interior degrees of freedom, which
 correspond to
gravitational interaction.   The topological description of graviton
 emission should be in
terms of the decay of  3-surface via topological sum vertex rather than the
decay of
 boundary
component as in case of gauge interactions.

\vm

The study of the   low energy field theory limit   \cite{lightfield}
defined as a  second quantized field theory inside
 p-adic
convergence
 cube  with side $L_p$ provides the practical manner to calculate at
energies below
 Planck
mass.  This field theory limit  is essentially the p-adic version of  the
 standard
model understood in perturbative sense since gauge group and the spectrum
of light
states apart from some
 exotic
states are identical.   The construction of this theory
 demonstrates how p-adicity solves the divergence problems
of
ordinary QFT. \\ i)  The construction of the momentum eigenstates involves
 elegant
 number
theory.   The discretization of momenta
 caused  by
the  finite size
of the
convergence cube implies automatically the absence of  infrared
divergences.\\
ii) p-Adicity implies automatically the absence of   ultraviolet
 divergences.
 Perturbation theory fails for certain critical size ($L_p$) of the p-adic
 convergence
cube   if self energy loops  contain massless
particles.   p-Adic planewaves actually
make possible to construct p-adic theory of integration and p-adic Fourier
analysis, which is a  central technical tool in the construction of the
theory.

\vm

The second  step in the program is the formulation of p-adic
 field theory limit
of TGD inside convergence cube  of size $L_p$ taking into account only
boundary degrees of freedom and treating boundary components as point like
objects but taking into account the  conformal degrees of freedom of
boundary
component somehow.
 The basic idea is to
generalize the p-adic counterpart of the standard model to a second
quantized
field theory using as basic fields Kac-Moody spinors rather than ordinary
finite component fields and to deduce basic vertices from the
requirements of
gauge and conformal invariance.

\vm

The third step would be the formulation of field theory description in
length scales
$L>L_p$. This theory is compeletely analogous to nonperturbative QCD
 describing hadrons
as basic objects.  An attractive possibility is that this theory can be
 constructed
 as second quantized theory in the discrete lattice
formed by p-adic convergence cubes (basic dynamical objects) and that
 vertices are
essentially the  n-point functions of the field theory inside single
 converge cube.

\vm

 The fourth step  would   be the  formulation of  quantum field
 description
 for gravitational interaction in terms of the interior degrees of freedom.
One should also construct a description for the topology changing
reactions   p-adic cubes describing for instance hadron reactions.

\vm

 During the work it has become clear
that time is perhaps  ripe for  an attempt to  formulate the general  p-adic
 field theory including both boundary and interior degrees of
freedom.  The construction could be  based on the following
basic ingredients.

\vm

\noindent
 a)   Field theory limit is defined as a purely fermionic QFT on quantum
averaged
spacetime defined as an absolute minimum of effective action: minimization
involves also
a variation in  the degrees of freedom associated with 3-surface  whereas
 in the
definition of K\"ahler function the absolute minimum is defined  for the
 variations
keeping the 3-surface  fixed.  Average spacetime depends on quantum state.
 The
enormous vacuum degeneracy of K\"ahler action (any surface, which is
submanifold of
$M^4\times Y^2$), where $Y^2$ is Lagrange submanifold of $CP_2$ having
vanishing
K\"ahler form) implies analogy with spin glass phase and suggests
ultrametricity of the
effective spacetime topology. A very general realization of ultrametricity
is provided
by the concept of topological condensate defined as a manifold obtained by
 gluing
together spacetime regions with different values of p-adic prime  $p$.
The hypothesis
about RG invariance of K\"ahler action  and symmetry arguments suggest
that effective
action  is  just the supersymmetrized p-adic counterpart of K\"ahler Dirac
action
defining configuration space geometry.\\ b)  Criticality suggests that
effective
action allows p-adic conformal transformations of $M^4$ as symmetries.
Actually
K\"ahler action allows all diffeormorphisms of $M^4$ as symmetries in the
excellent
approximation that induced metric is flat.   Also the canonical
transformations of
$CP_2$ coordinates localized with respect to $M^4$ coordinates are
symmetries in same
approximation in the  interior of $X^4$.   Both symmetries become
 exact if
one requires
   that the infinitesimal generators of these transformations
  annihilate  physical states. \\  c) $ N=1$ supersymmetry generated
by right handed neutrino  is realized  if one assumes that
quantized $H$ coordinates contains super part. In $CP_2$ degrees of freedom
super symmetrization is defined in preferred $CP_2$  coordinates, for which
 the action
of  $su(2)\times u(1)$ is linear. These coordinates are determined up to a
 color
rotation
  and internal consistency requires that physical states are color singlets
 (each
p-adic region of spacetime is color singlet).   This implies that $H$
 coordinate
becomes dynamical quantum field and gauge bosons and  gravitons
 correspond to 2-   and
 4-fermion
bound states  respectively.  $N=1$ supersymmetry and
closely related  Kappa symmetry make it possible to extend conformal
symmetry  to
super conformal symmetry on boundaries.
$Diff(M^4)$ and canonical transformations have similar extension in the
interior of
$X^4$.  \\  d) In the interior of $X^4$  the quantization in fermionic
degrees of freedom relies on the properties of p-adic planewaves.
 $Diff(M^4)$
 and local canonical invariance pose additional constraints on the
states.     $Diff(M^4)$ invariance allows massless boson
states annihilated by all $Diff(M^4)$ generators except translations.
Graviton is contained
in the spectrum but many fermion states are not possible in accordance with
 the basic
assumption that partons correspond to boundary components.  p-Adic
perturbation theory
is free from UV divergences but infrared cutoff is necessarily present.\\
 e) The
practical treatment of boundary components involves idealization to world
 line and the
use of Kac Moody spinors to describe the vibrational degrees of freedom
 of boundary
component.  The assumption that  Super Virasoro generators are sums of
Super Virasoro
generators associated with boundary component time coordinate and with
 2-dimensional
boundary component explains  Kac Moody Dirac equation used in the
 description of Higgs
mechanism.  Conformal $so(3,1)$, $so(4)$ and $su(3)$ are approximate
 symmetries in
boundary degrees of freedom.  Second quantization must be performed
 for  Kac
Moody spinors.

\vm

Although this rough general formulation need not yet offer practical
calculation
 receipes and need not even be correct, it makes possible to understand p-adic
topology   as a property of quantum averaged spacetime and demonstrates that
p-adic
 super
conformal invariance on boundaries, crucial for p-adic description of Higgs
mechanism, is not in conflict with the basic principles of
TGD.    In principle the formulation  offers a possibility to derive vertices
for  simpler
p-adic conformal field theory limit formulated in terms of Kac Moody
spinors
only and treating boundary components as point like objects.  If
the general formulation is correct then the main
unsolved  technical problem is the construction of  the interaction vertices
 for N-S and
Ramond typer Kac Moody spinors.

\vm

The program of the work is following: \\
a) The general ideas of Quantum TGD are described.\\
b) p-Adic building blocks of the theory  are described.
\\ c) The concept of effective action is defined and the p-adization of
 spacetime
in terms of topological condensate concept is described. \\
d)  The symmetries of the theory are discussed.
 \\ e) A formulation of  field theory limit in terms of p-adicized
K\"ahler Dirac action in the interior of $X^4$ is disccused in  detail. \\
f) Appendix contains general
formulas for  currents associated with symmetries.

\section{Quantum TGD and p-adic field theory limit}

In the following some basic facts about TGD are considered and shown to lead
 to quite unique  general receipe for the construction of Quantum TGD.

\subsection{Construction of configuration space metric and spinor structure}

 The construction of Quantum TGD reduces to the construction of geometry for
 configuration space consisting of all possible 3-surfaces in space
$H=M^4_+\times CP_2$.   What is needed is  metric and spinor structure.
The construction is based on the assumption that the geometry in question is
K\"ahler geometry.  This means that configuration space allows complex
structure
and metric can be expressed in terms of K\"ahler  function.  A further
requirement is $Diff(X^4)$ invariance and this requires that the definition
of
K\"ahler function associates to a 3-surface 4-surface at which $Diff^4$ can
 act.

\vm

The value of
K\"ahler function  for a given 3-surface  $X^3$  is defined as absolute
minimum
 of K\"ahler action in the set of 4-surfaces having $X^3$ as submanifold is
 defined as Maxwell action for Maxwell field defined by induced K\"ahler
 form of
$CP_2$. Action contains also Chern-Simons term associated with boundaries of
4-surface.  Definition associates to a given 3-surface the absolute minimum
spacetime surface identifiable as 'classical history'. One interesting
feature
of the  construction is that K\"ahler function defines generalized
catastrophe
 theory
in configuration space  (studied in the third part of the book).   K\"ahler
action is analogous to
 thermodynamic free
energy and  can have several extrema and absolute minimum requirement
selects
one of the extrema just as minimization of free energy selects the
thermodynamic
phase.   The properties of absolute minimum 4-surface can change
discontinously
along  'Maxwell line' just as thermodynamic phase transition occurs along
Maxwell line.

 \vm

The definition of spinor structure necessitates the definition of
anticommuting gamma matrices. These are defined  for a given $X^3$ as linear
superpositions of fermionic oscillator operators associated with second
quantized Dirac action for induced spinors in the classical space time
 surface
$X^4(X^3)$.   The action is just free Dirac action on background defined by
induced spinor connection in $X^4(X^3)$.

\vm

The proposed  more explicit construction of configuration space metric is
 carried on the set of 3-surfaces belonging to  'lightcone boundary'
 $\delta
M^4_+ \times CP_2$.  The motivation is that $Diff(X^4)$ invariance dictates
 the
metric from its initial values on light cone boundary.  The generalized
 conformal
structure of the light cone boundary having metric dimension $2$ plus the
assumption that metric allows infinite-dimensional Kac Moody symmetry
 (Lie
algebra corresponds to the group of maps from $\delta M^4_+$ to $su(3)$)
 plays
essential role in the construction.
 The conformal structure of light cone boundary and  Kac Moody algebra
structure should have some counterparts in the construction of conformal
field
field limit. It is to be emphasized that the construction of configuration
 space
geometry and spinor structure is by no means a closed chapter of TGD and
the support
for the essential correctness of the picture comes from various applications
 rather
than from rigorous mathematical formulation.

\subsection{Some basic properties of K\"ahler action}

\subsubsection{ Criticality of the  K\"ahler action}

Quantum TGD should define a theory of everything.  For instance,  gauge
couplings
should be predictions of the theory.    K\"ahler action contains K\"ahler
coupling strength $\alpha_K$ as free parameter and one should have some
principle to
determine the value of this parameter.

\vm

The  key observation is that the vacuum functional of the theory is
 uniquely
 defined as  exponential of K\"ahler function from the  requirement that
 two ill
defined determinants, namely the determinantof the configuration space
metric and the
Gaussian determinant associated with perturbative functional integration
 around
minimum of K\"ahler function cancel each other.  The vacuum functional is
analogous to thermodynamic partition function and K\"ahler coupling strength
 is
analogous to temperature.  Therefore a natural requirement is that the value
 of
K\"ahler coupling strength $\alpha_K$  is such that Quantum TGD describes
 critical
system.

 \vm

The duality of $CP_2$ K\"ahler form supports the criticality hypothesis: in
standard complex coordinates  of $CP_2$ the
geometric realization of duality tranformation is as an orientation
changing
isometry $\xi_1\leftrightarrow \xi_2$. In certain $N=1$ supersymmetric
theories duality transformation relates strong coupling phase for particles
with weak coupling phase for magnetic monopoles. In self dual case coupling
constant and action must be critical and particles and magnetic monopoles must
be equivalent.  The so called $CP_2$ type extremals \cite{TGD} identified as
elementary particles can indeed be regarded as magnetic monopoles in
homological sense.  What is remarkable is that elementary particles
create  no long range
$1/r^2$ type magnetic field  since magnetic flux flows in $CP_2$
degrees of freedom: this is possible by to the nontrivial homology
 of $CP_2$. Also magnetic monopoles in ordinary sense are in principle
possible but are  not favoured
by the minimization of K\"ahler action: magnetic $1/r^2$ field gives
positive contribution to the action whereas electric $1/r^2$ field
gives negative contribution and is  therefore favoured.

\subsubsection{Coupling constant evolution hypothesis}

Coupling constant evolution hypothesis is based on the assumption that
 K\"ahler coupling strength is a fixed point of coupling constant evolution
 and on
the observation that K\"ahler action can be interpreted as EYM action for

induced gauge fields (electroweak gauge fields and gluons) and induced
 metric.
Besides this one must include the field defined by induced gamma matrices
 $\Gamma_{\alpha}= \Gamma_k \partial_{\alpha}$, which
was identified as a field describing the presence of fermions in \cite{TGD}.
 It
seems that this interpretation is not correct but that new boson field is in
question.   The hypothesis gives nontrivial constraints between various
coupling
strengths in accordance with general features of the  coupling constant
 evolution and
p-adic  field theory limit should provide quantum level realization for
the coupling constant evolution hypothesis.
The criticality is expected to fix the value of K\"ahler coupling strength
 as a
quantity analogous to critical temperature so that in principle all coupling
constants come out as  predictions.

\subsubsection{Vacuum degeneracy of K\"ahler action and canonical
 invariance}

K\"ahler action allows enormous vacuum degeneracy: any 4-surface with
$CP_2$
projection, which is Lagrange manifold is vacuum extremal. Lagrange
manifold
is defined as a submanifold of $CP_2$ with vanishing induced  K\"ahler
 form with
dimension  $d\leq 2$.  Canonical transformations leave K\"ahler form
invariant and
are obviously exact dynamical symmetries in vacuum sector.

 \vm

The vacuum degeneracy implies that the second variation of K\"ahler action,
whose inverse would define propagator in  the usual quantization, is highly
degenerate.  In fact, for the standard imbedding of  $M^4$    the lowest
 order
term in action is fourth order in the deviation of  $CP_2$ coordinates from
constant values
so that canonical quantization around flat solution is not possible.   In
 Quantum
TGD this kind of quantization is not needed. If field theory limit is
 defined as a
field theory defined on quantum averaged spacetime  defined by effective
action this
problem is  not  encountered    Since $CP_2$ coordinates are treated
classically they
define external fields for the  quantum theory formulated in terms of
fermionic fields
only. This is in accordance with the result of Quantum TGD  that boson
 quanta
correspond fermion-antifermion bound states on boundaries. The existence
of long
range classical gauge boson fields, in particular $Z^0$ fields,
have many applications in TGD \cite{TGD} so that the   appeareance of
these fields in the formulation of quantum theory is a rather satisfactory
feature.

\vm

The vacuum degeneracy makes  TGD:eish Universe analogous to spin glass phase.
Fermions correspond to spins, induced gauge potentials and induced metric
correspond to couplings between spins and the presence of vacuum degeneracy
implies that one must take average over a large number of vacuum
 configurations
with different gauge fields.  Induced K\"ahler form is analogous to the
frustration field. Ultrametricity is the basic property of the
 configuration space
for the minima of effective action in spin glass phase. This suggests that
the
effective spacetime topology associated with effective action is ultrametric.
p-Adic ultrametricy is especially attractive since it makes possible  p-adic
 conformal
invariance.  The most general form of p-adic topology is obtained by gluing
together
p-adic regions of spacetime with different p-adic $p$ along their common
 boundaries in
a manner to be described later.  Topological condensate is p-adic manifold
$..<p_1<p_2...$ consisting of p-adic surfaces glued together along common
 boundaries
with the property that the sizes of the surfaces form increasing hierarchy.

\section{p-Adic building blocks }

The construction of p-adic  field theory limit necessitates the
generalization of the  basic tools of standard physics  such as differential
 and
integral calculus, the concept of Hilbert space, Riemannian geometry, group
theory, action principles, probability and unitary concepts  to  p-adic
context. The construction of all these tools relies heavily
 on the  canonical identification of p-adic and real numbers.

 \subsection{Canonical identification}

The canonical identification
 $x=\sum_m x_mp^n \in R_p \rightarrow  \sum x_np^{-n} \equiv x_R \in R $ of
p-adic and real numbers is  the cornerstone of p-adic TGD.  The success of
 the
elementary particle mass calculations and the possibility to understand
connection between p-adic and real probabilities suggests that canonical
identification is related to  some deeper mathematical structure.  The
following inequalities hold true:

\begin{eqnarray}
(x+y)_R&\leq& x_R+y_R\nonumber\\
\vert x\vert_p \leq (xy)_R&\leq &x_Ry_R
\end{eqnarray}

\noindent where  $\vert x \vert_p$ denotes p-adic norm.  These inequalities
 can be generalized to case of $(R_p)^n $ ( linear space over
p-adic numbers).

\begin{eqnarray}
(x+y)_R&\leq& x_R+y_R\nonumber\\
\vert \lambda vert_p \leq (\lambda y)_R&\leq &\lambda_Ry_R
\end{eqnarray}

\noindent where the norm of the vector $x\in T_p^n$ is defined in some
 manner.
 The case of Euclidian space suggests the definition

\begin{eqnarray}
(x_R)^2 &=& (\sum_n x_n^2 )_R
\end{eqnarray}

\noindent   These inequilities resemble those satisfied by vector norm.
The only difference is  the failure of  the linearity in the sense that the
 norm of
scaled vector is not obtained by scaling the norm of original vector.
 Ordinary
situation prevails  only   if scaling corresponds to power  of $p$.

These observations suggests that the concept of normed space or Banach
space
might have generalization and physically the generalization might apply to
 the
description of highly nonlinear systems such as TGD. The nonlinearity would
 be
concentrated in the nonlinear behaviour of the norm under scaling.

\vm

  Amusingly,
the p-adic counterpart of Minkowskian norm

\begin{eqnarray}
(x_R)^2 &=& (\sum_k x_k^2- \sum_l x_l^2 )_R
\end{eqnarray}

 \noindent  produces nonnegative  norm.
Clearly the p-adic space with this norm  is analogous to future light cone.

\vm

\subsection{  Existence of square roots of p-adically real numbers and 4- and
8-dimensional extensions of p-adic numbers}.

The   existence of square root plays key role in various p-adic
 considerations.
\\
a) The p-adic generalization of the representation theory of the
 ordinary groups and  Super Kac Moody and Super Virasoro algebras
 exists provided an extension of p-adic numbers allowing square roots of
 p-adically
 real numbers  is used.
 The reason is that the  matrix elements of the raising and lowering
 operators as well
 as oscillator operators typically involve square roots.  Also the
 definition of
 anticommuting Gamma matrix fields in the representations of Super Virasoro
 algebra
 involves square root
 function  $\sqrt{Z}$ \cite{Conformal} of algebraically extended p-adic
 number. \\
b) The existence of square root is  a  necessary ingredient in the
 definition of p-adic unitarity and quantum probability  concepts. \\
c) p-adic Riemannian geometry  necessitates the existence of square root
since the
definition of the infinitesimal length $ds$ involves square root. \\
 For  $p>2$ the minimal dimension for  the algebraic extension allowing
square roots
 is $D=4$.  For $p=2 $ the dimension of the minimal extension is $D=8$.
This implies
that the  dimension of the  p-adic Riemannian geometries considered in
the first part
of the paper is multiple of $4$ or $8$ depending on whether $p>2$ or $p=2$
 so that
p-adic Riemannian geometry gives very deep explanation for  spacetime
 and imbedding space dimensions.

\vm

In $p>2$ case the general form of extension is

\begin{eqnarray}
Z&=& (x+\theta y) +\sqrt {p}( u +\theta v)
\end{eqnarray}

\noindent  where the condition $\theta^2= x$ for some p-adic  number $x$
 not allowing
square root as a p-adic number.  For $p \ mod \ 4 =3$ $\theta$ can be taken
 to be
imaginary unit. This extension is natural for p-adication of spacetime
 surface so that
spacetime can be regarded as a number field locally.  Imbedding  space can
 be regarded
as  a cartesian product of two 4-dimensional
 extensions locally.

 \vm

 In $p=2$ case 8-dimensional extension is needed to define square roots.
 The extension is defined by adding  $\theta_1= \sqrt{-1}\equiv i$,
$\theta_2=
 \sqrt{2}$, $\theta_3= \sqrt{3}$ and the products of these so that
the extension can be
written in the form

\begin{eqnarray}
Z&=& x_0+ \sum_k x_k\theta_k + \sum_{k<l} x_{kl} \theta_{kl}+
x_{123}\theta_1\theta_2\theta_3
\end{eqnarray}

 \noindent Clearly,  $p=2$ case is exceptional as far as the  construction
of the
 conformal
 field theory limit is considered since the structure of the  representations
 of Virasoro algebra and groups in general changes drastically in $p=2$ case.
  The
 result suggest that
 in $p=2$ limit spacetime surface and $H$ are in same relation as real
 numbers and complex numbers: spacetime surfaces defined as the absolute
 minima of 2-adiced K\"ahler action are perhaps identifiable as  surfaces
 for which the imaginary part of 2-adically analytic function in
$H$ vanishes.

\vm

The physically interesting feature of p-adic group representations is that
if
one
doesn't use $\sqrt{p}$ in the extension the number of allowed spins for
 representations
of $SU(2)$ is finite: only spins $j<p$ are allowed. In $p=3$ case just
 the spins $j\le
2$ are possible. If 4-dimensional extension is used for $p=2$ rather than
 8-dimensional
then one gets the same restriction for allowed spins.

\subsection{p-Adic spacetime and topological condensate as generalized
 manifold}

p-Adic topology and differentiability rather than ordinary topology and
differentiability are useful concepts above some length scale $L_0$ of
the
order of $10^4$ Planck lengths as suggested by the mass scale argument.
 The
p-adic realization of the Slaving Principle suggests that  various levels
of the topological condensate correspond to different values of p-adic
 prime $p$: the
larger  the length scale the larger the value of $p$ and the course the
 induced real
topology.  If the most interesting  values of $p$ indeed correspond
 Mersenne primes
the number of most interesting levels  is  finite: at most $12$ levels
 below electron
length scale: actually also  primes near prime powers of two seem to be
 physically
important.    According to
the experiences achieved from  the study low energy field theory limit
 and
contrary to the original expectations,  $L_p= \sqrt{p} L_0$  serves as
 infrared cutoff
rather than ultraviolet cutoff  for the
 p-adic field theory. If length scale resolution is larger than $L_p$
ordinary real
physics should be an excellent approximation.

\vm

The existence of p-adic definite integral  (relying on the
canonical identification between p-adic and real numbers) and of  p-adic
valued inner
product
makes possible to  define the concept of Riemannian
metric and related structures  on spacetime surface and  the concept of
 submanifold
 with
induced  metric and  spinor structure makes sense.  Basic symmetry groups
 (in
particular, Poincare group)  acting as isometries of the imbedding space
 have p-adic
counter parts.   It is  possible   define variational principles for the
fields defined
in p-adic spacetime using p-adization of ordinary real action principle
and the
minimization of  the  K\"ahler action leads to the p-adic counterparts of
field
equations having same form as in real case.

\vm

The p-adic realization of the topological condensate concept requires
 rather general manifold concept. This manifold  has locally p-adic topology
 but
decomposes into regions with different values of p-adic prime $p$.   These
different
regions are glued together along their boundaries.   The
mathematical description of the  gluing operation for two p-adic topologies
 with
 different
values of $p$ is based on the use of canonical identification map.  What
 is
involved is the identification map identifying the boundaries of regions
with
topologies $p_1$ and $p_2$.   A possible manner to perform this
identification is  via
two steps:\\
a) The points $x_{p_1}$ of $R^n_{p_1}$ (region of boundary is subset of
 $R^n_{p-1}$)
are mapped via canonical identification to the points of  $R^n$:
 $ x_{p_1}\rightarrow
(x_{p_1})_R$.\\ b) These points of $R^n$ are expressed using $p_2$-adic
 pinary
expansion  and mapped via the
inverse of the canonical identification map to $R^n_{p_2}$, which is two
 valued for
real numbers with finite number of $p_2$-nary digits.

\vm

A more general but essentially equivalent  manner to describe gluing
 operation
is to
map
the common boundaries of $p_1$- and $p_2$-adic regions to regions of
 $R^n$ and to
identify these two regions of $R^n$  with differentiable invertible map.
 The possible difficulties in
the construction are related to the nonuniqueness of the inverse of the
 canonical
identification map implying that real  numbers with finite number of
 pinary digits
have two inverse images.

\vm

The construction generalizes to the  gluing  together  p-adic  submanifolds
of
$H$. For $p_1$ and $p_2$ imbedding space is regarded as $p_1$ and $p_2$-adic
 manifold
respectively. What is required is that the real counterparts for the
imbeddings space
points associated with identified  boundary points  are identical. This
means that the
condition

\begin{eqnarray}
(h^k(x_{p_1}))_R&=&(h^k(x_{p_2}))_R \  \ for \ \ (x_{p_1})_R= (x_{p_2})_R
\end{eqnarray}

\noindent holds true. For large values of $p$  already the lowest pinary
digit
gives an excellent discretization of the  p-adic convergence cube and in
this
approximation the conditions defining gluing are rather simple
numerically.
Analytically the gluing conditions do  not look tractable.

\subsection{p-Adic unitarity and probability concepts}

 p-Adic unitarity and probability
concepts  lead  to highly nontrivial conclusions
concerning the general structure of the S-matrix.  S-matrix can be expressed
as

\begin{eqnarray}
S&=& 1+i\sqrt{p} T\nonumber\\
T&=&O(p^0)
\end{eqnarray}

\noindent for $p \ mod \ 4=3$ allowing imaginary unit in its
 four-dimensional
 algebraic extension. Using the form $S= 1+iT$, $T= O(p^0)$  one
 would obtain in
general transition rates of order inverse of Planck mass and theory would
have
nothing to do with reality.   Unitarity requirement implies iterative
expansion
of $T$ in powers of $p$ (\cite{lightfield} and  the few
lowest powers of $p$ give extremely
 good
approximation for the  physically most interesting values of $p$.

\vm

 The  relationship between  p-adic and real probabilities involves   the
hypothesis \cite{lightfield} that
transition probabilities depend on the experimental resolution.
Experimental resolution is defined by the decomposition of the  state space
$H$  into direct sum $H=\oplus H_i$ so that experimental situation cannot
differentiate between different states inside $H_i$.   Each resolution
defines different real transition probabilities unlike in
ordinary quantum mechanics.  Physically this means that   the experimental
arrangements,  where one monitors each state in $H_i$ separately differ from
the situation,  when one only looks whether the state belongs to $H_i$.
One  application is related to the  momentum space resolution dependence of
the
transition probabilities. More exotic application \cite{lightfield} is
related to  $Z^0$  decay widths: the total annihilation rate to excotic lepton
pairs (unmonitored) is essentially zero:
 if one
would (could)  monitor each excotic lepton pair one would obtain simply sum
of the
rates to each pair.

\subsection{p-Adic thermodynamics and transition probabilities}

The p-adic field theory limit as such is not expected to give
 a realistic theory at elementary particle physics level.  The point is that
particles are expected to be either massless  or possess Planck masses.
 The
 p-adic description of
Higgs mechanism documented  shows that p-adic thermodynamics provides the
proper formulation of
 the
problem.   What is thermalized is Virasoro generator $L_0$ (mass squared
contribution is not included to $L_0$ so that states do not have fixed
 conformal
weight).   Temperature is quantized purely number theoretically in low
temperature limit ($exp(H/kT)\rightarrow p^{L_0/T}$, $T=1/n$): in fact,
 partition
function does not even exist in high temperature phase.  The extremely
 small mixing of
massless states with Planck mass states implies massivation and predictions
for
elementary particle masses are in excellent agreement with experimental
 masses.
Thermodynamic approach also explains the emergence of  the length scale
$L_p$ for
a given
p-adic condensation level and one can  develop arguments explaining why
 primes
near prime powers of two   are favoured.

\vm

In the calculation of S-matrix thermodynamics approach should work also.
 What is is needed is thermodynamical expectation value for transition
amplitudes squared over incoming and outgoing states. In this expectation
value
3-momenta are fixed and only mass squared varies.

\subsection{p-Adic cancellation of UV divergences}

p-Adic cancellation of the  ultraviolet divergences is based on very simple
observations. The necessary infrared cutoff $L_p$ implies infrared cutoff
 and
discretization for momentum: what one has is just quantum field theory in
finite
4-dimensional box.  Loop integrals over the  loop momenta are replaced by
summation over momenta and for a given value of the propagator expression
one always
gets integer valued degeneracy factor $N$ giving the   number of momentum
combinations giving  rise to this particular value of the propagator
expression.
Since integer has always p-adic norm not larger than one this means that
the
summations over the loop momenta converge trivially provided the integers
 $N$
are finite.

\vm

  For standard $M^4$ metric it was found that for spacelike propagator
momentum $P_{in}$  the degeneracy factor $N=N(P_{in},k)$ is for certain
values of
propagator momentum $k^2$ infinite as real number and although the p-adic
 norm of
$N(P_{in},k)$ is not larger than one it is ill defined p-adically.   It
 was
found that the gravitational warping of the   standard  imbedding of $M^4$
($CP_2$
coordinates cease to be constant and $CP_2$ projection of spacetime
 becomes
geodesic circle)  to flat nonstandard imbedding transforms the induced
metric
$m_{\alpha\beta}$ to
$  m_{\alpha\beta}- p^2KR^2 P_{\alpha}P_{\beta}$,   where $P$ is the total
momentum associated with the  p-adic convergence cube. Replacing the
standard inner
product for momenta with the  inner product defined by the  warped metric
 the
degeneracies associated with  the p-adic propagator expression become
finite.

\vm

 The  necessity to take into account the gravitational warping of the flat
spacetime
means that one cannot totally neglect  the action of the fermionic  matter
 on the
background spacetime geometry.  The
back-reactions related to other conserved charges are also expected to be
present but one might hope that these effects  are small.
In principle one can take into account back reactions exactly by
interpreting field equations as expectation values between various states.
 In
particular, $H$-coordinates can depend on various conserved charges
 besides
four momentum, boundary components serve as sources of classical gauge
fields,
etc..

\subsection{p-Adic Fourier analysis}

The definition of p-Adic integration is based on two ideas. The ordering for
 the limits of
integration is defined using canonical correspondence. $x<y$ holds true if
 $x_R<y_R$ holds true.
The integral functions can be  defined for Taylor series expansion by
 defining
indefinite integral as the   inverse of the differentiation.  As far as
 physical
 applications is
considered  this definition does not seem to be the correct one since p-adic
plane waves are
not analytic functions. p-Adic planewaves are certainly needed since ordinary
exponential
 functions $exp(ikx)$ have unphysical properties and orthonormalization does
not have seem any obvious physical meaning. For p-adic planewaves
orthogonalization is natural  in light  of  their  periodicity properties.
Therefore in the sequel all functions are assumed to be expressible as
 p-adic
Fourier series using p-adic plane wave  basis whose detailed definition
is given
in \cite{lightfield}. In the region
$0\leq x_R\leq (p-1)$
the basis functions are given by

\begin{eqnarray}
f^k(x)&=& a^{kx}, k=0,...,p-1\nonumber\\
a^{p-1}&=&1
\end{eqnarray}

\noindent In smaller regions $ 0\leq (x-x_n)_R \leq (p-1)/p$ the same
 functional
form applies expect that p-adic $k$ is scaled to $k/p$. One can associate
with the cm coordinate  $x_n=n$ of this region  p-adic planewave with
 momentum so
that
one obtains planewaves with momenta $m/p +k$, $m,k =0,...,p-1$.   This
 process
of going to smaller and smaller  length scales can be continued
indefinitely
and one obtains p-adic planewaves with infrared cutoff in momentum.
Momentum
eigenstate basis  has as its real analog  so called wavelet  basis
\cite{Wavelets}: this
basis has been found to be especially effective in the numerical treatment
 of wave
equations.

\vm

p-Adic planewaves satisfy periodic boundary conditions
 $f_k(0)= f_k(p-1)=1$.
One can define also p-adic planewaves satisfying antiperiodic boundary
conditions as square roots of p-adic planewaves for odd values of momentum
 $k$.
The square root indeed exists for the extension allowin square root of real
p-adic number.
These two kinds of planewave basis are essential for the realization of
 Ramond
and N-S type representations of Super Virasoro.

\vm

 The orthonormalization of
p-adic planewaves gives definition of p-adic integration in the range
$(0,p-1)$.
All other p-adic planewvaves except constant  $k=0$ planewave have vanishing
integral. Constant planewave gives integral, which can be chosen to be some
constant. This definition of integration is of central importance in the
formulation of variational principles based on Lagrangian density.

\vm

Although p-adic planewaves are real they allow complex conjugation type
operation  defined by $k\rightarrow -k$.   p-Adic planewaves possess infrared
cutoff and this means that p-adic momentum $k$ can have only a finite
number of
non-nonnegative pinary digits whereas $-k=k(p-1)(1+p+p^2+....) $ has
infinite
number of pinary digits.  Therefore only positive values of $k$ appear in
the
Fourier expansion of any function $f$ and the conjugate of $f$ appearing
 in
the definition of  the inner product do not belong to the Hilbert space in
 question
so that Hilbert space and its dual are not identical for p-adic planewaves.
For higher dimensional p-adic planewaves all momentum components have
 finite number of pinary digits and this implies some new features in
quantization. For instance, for fermion field  positive energy modes
creating
fermions have p-adic momenta have $k_i>0$ for each $i$ whereas negative
energy modes destructing antiferions have p-adic momenta $k_i<0$ for
 each $i$.
The generalization of positive/negative energy concept to all momentum
components makes possible to realize approximate
$Diff(M^4)$  invariance in a  nontrivial manner.

\vm

There is problem related to the definition of derivative for p-adic
 planewave.
p-adic planewave is not differentiable in ordinary p-adic sense. The analogy
with real planewaves

\begin{eqnarray}
a^{kx}&"="& exp(\frac{i2\pi k x}{(p-1)}) \  "=" ln(a)k
\end{eqnarray}

\noindent suggests

\begin{eqnarray}
da^{kx}/dx&"="& ln(a)k a^{kx} \  "="  \frac{i2\pi k}{(p-1)}a^{kx}
\end{eqnarray}

\noindent  which however does not
exist p-adically. There are several manners to avoid the problem. Consider
first the tricky manners:\\
  i)  Define $\pi$ as the p-adic counterpart of real
$\pi$ obtained via  the canonical identification. The presence of the
imaginary unit
is troublesome since p-adic planewaves are real in the ordinary sense. \\
ii) Define $ln(a)$ as the p-adic counterpart of $ln(a_R)$ defined by
 the canonical
identification. In this case one avoids imaginary units.

\vm

Neither of these  tricks is needed in the  formulation of the  p-adic field
theory
limit of TGD  since  the  action is well defined apart from a
proportionality factor $ln(a)^{-D}$, $D=4$ in interior and $D=3$ on
 boundaries
and by defining the interaction measure $dx$ to be proportional to
 $1/ln(a)$ for
each coordinate $x^k$ the action becomes independent of $ln(a)$. This
 is seen
by following arguments: \\
a) For surfaces representable as  maps from $M^4$ to $CP_2$ (the
restriction is not essential) $CP_2$ coordinates must allow p-adic Fourier
expansion and therefore $CP_2$ part of the induced metric is necessarily
proportional to $ln(a)^2$.  The same must be true for $M^4$ part of the
 induced
metric and the imbedding is of the  following general form

\begin{eqnarray}
m^k &=& ln(a)x^k+ Fourier \ expansion\nonumber\\
s^k&=& Fourier \ expansion
\end{eqnarray}

 \noindent The form of $M^4$ coordinate does not lead to  difficulties
since metric of $H$ does not depend on $M^4$ coordinates and  $m^k$ appears
 only
in p-adic planewaves $exp(p_km^k)=a^{kp_kx^k}$. In $M^4$ coordinates p-adic
planewaves are formally similar to ordinary planewaves. \\
b)  Also the super
symmetrization of K\"ahler action to obtain effective action  is possible
 by the
replacement $h^k \rightarrow h^k_S\equiv h^k + \bar{\Psi}\gamma^k\Psi$.
  The
supersymmetrized planewaves $exp(m^k_Sp_k)$ are well defined

\begin{eqnarray}
exp(p_km^k_S)&=&  a^{p_kx^k}exp(p_k\bar{\Psi}\gamma^k\Psi)
\end{eqnarray}

\noindent b) Induced spinor connection is proportional to $ln(a)$ and in
 Dirac
operator $ln(a)$ dependences coming from induced gamma-matrices and
covariant
derivative cancel each other.  K\"ahler  Dirac action density is
proportional to
$ln(a)^D$, $D=4(3)$ in interior (boundary) and by defining  coordinate
differential $dx^k$ so that it contains $ln(a)$ factor one obtains
 completely
well defined action density.

\section{Length scale hypothesis}

p-Adic length scale hypothesis states that the
length scales $L_p=\sqrt{p}L_0$  associated with primes near prime powers
of 2,
especially Mersenne primes are physically especially interesting.
The results of  the  mass calculations  give convincing
support
for the  hypothesis and explain the emergence of $L_p$ for a given $p$ in
 terms of
p-adic thermodynamics for Super Virasoro representations. There are also
many
arguments giving partial explanation for Mersenne primes and primes
near prime powers of two.  In  the following one of the arguments for
 believing that
primes near prime powers of two are physically most interesting is produced
 and the physical interpretation of $L_p$ and
$L_0$  is considered.

\subsection{ Length scales defined by
 prime powers of two and Finite Fields}

Mersenne primes ought to be physically especially interesting primes by
arguments relying on 2-adic fractality.  Above  $M_{127}$  there is
extremely
large gap for Mersenne primes
 and this suggests that there must be also other physically important
primes.  Certainly all primes near powers of $2$ define physically
interesting length scales by 2-adic fractality but there are two
 many of them.  The first thing,  which comes into mind is to consider  the
set of   primes near prime  powers of two  containing as special case
Mersenne primes.
One can indeed  find a good argument in favour of these length scales, which
play central role in the p-adic description of Higgs mechanism.

\vm

TGD Universe is critical at quantum level and criticality is  related
 closely to scaling invariance.
This  suggests that    unitary irreducible representations of p-adic
scalings $x \rightarrow p^mx$, $m \in Z$ should play central role in
 quantum theory.  Unitarity
requires that  scalings are represented by a multiplication with phase
factor and
 the reduction to a representation of finite cyclic group $Z_m$ requires
 that
 scalings $x\rightarrow
p^mx$,
 $m$ some integer,  act trivially.  In ordinary complex  case the
 representations
 in question
correspond to phase factors
  $\Psi_k(x)= \vert x\vert ^{(\frac{ik2\pi}{ln(p)})}=
exp(iln(\vert x\vert)\frac{k2\pi}{ln(p)})$,  $k\in Z$
and the reduction to a representation of $Z_m$ is also  possible but there
 is
no good reason for restricting the consideration to discrete scalings.

\vm

a)   The Schr\"odinger amplitudes in question
 are p-adic counterparts of ordinary complex functions
 $\Psi_k(x)= exp(iln(\vert x\vert)k\frac{ik2\pi}{ln(p)})$, $k\in Z$.  They
 have   unit
 p-adic norm, they are  analogous to planewaves,  they depend on p-adic norm
 only
 and satisfy the
scaling invariance condition

\begin{eqnarray}
\Psi_k(p^mx\vert p\rightarrow p_1 )&=&\Psi_k(x\vert p\rightarrow p_1)
\nonumber\\
\Psi_k(x\vert p\rightarrow p_1)&=& \Psi_k(\vert x\vert_p\vert
 p\rightarrow p_1)\nonumber\\
\vert \Psi_k (x\vert p\rightarrow p_1)\vert_p &=& 1
\end{eqnarray}

\noindent which guarantees that these functions are effectively functions
on p-adic
numbers
with cutoff performed in $m$:th power.   \\
b) The solution to the conditions is suggested by the analogy with the real
 case:

\begin{eqnarray}
\Psi_k(x\vert p\rightarrow p_1)&=& exp(i\frac{kn(x)2\pi}{m})\nonumber\\
n(x)&=& ln_p(N(x)) \in N
\end{eqnarray}

\noindent where $n(x)$ is integer (the exponent of the lowest power of
 p-adic number)
and  $k=0,1,...,m-1$ is integer.  The
existence of the functions is  however not obvious.  It will be shortly
  found that
the functions  in question  exist in  $p>2$-adic for all $m$ relatively
 prime with
respect to $p$ but
exist for all odd $m$  and $m=2$ in 2-adic case. \\
c) If $m$ is prime (!)  the functions
$K=\Psi_k$
form a finite field $G(m,1)= Z_m$
 with respect to p-adic sum  defined as the p-adic product
 of the  Schr\"odinger amplitudes

\begin{eqnarray}
K+L &=& \Psi_{k+l}= \Psi_k\Psi_l
\end{eqnarray}

\noindent and multiplication defined as

\begin{eqnarray}
KL&=& \Psi_{kl}
\end{eqnarray}

\noindent So,  if the proposed Sch\"odinger amplitudes possessing
 definite scaling invariance properties are physically important
 then the length scales defined by the  prime powers of two
 must be physically special since Schr\"odinger amplitudes  or equivalently,
 the p-adic scaling momenta $k$ labeling them,
  have natural finite field structure. By Slaving Hierarchy Hypothesis also
the p-adic length scales near prime powers of two (and perhaps of prime
 $p>2$, too)
are
therefore physically interesting.

\vm

  As shown in \cite{Mathideas}, for  $p>2$  only those finite fields
 $G(p_1,1)$
for which $p_1$ is factor of $p-1$ are realizable as representation of
 phasefactors whereas
for $p=2$ all $G(p_1,1)$ allow this kind of representation.   Therefore
 $p=2$-adic
numbers are clearly exceptional.  In p-adic
 case the
 functions $\Psi_p(x,\vert p\rightarrow p_1)$  give   irreducible
representations
for the group of p-adic scalings $x\rightarrow p^mx$, $m\in Z$ and the
 integers $k$
can be regarded
as scaling momenta.  This suggests that these functions should play the
  role of
the  ordinary
momentum eigenstates in the  quantum theory of
fractal structures.
 The result motivates the hypothesis that   prime powers of $2$ and
 also of
$p$ define physically
 especially interesting p-adic length scales:  this hypothesis  will be
 of utmost
 importance in
future applications of TGD.

\vm

The  ordinary p-adic planewaves associated with translations can be
constructed as
functions $f_k(x)= a^{kx}$, $k=0,...,n$, $a^{n} =1$. For $p>2$ these
 planewaves are
periodic with period $n$,   which is factor of $p-1$ so that wavelengths
correspond to factors of $p-1$ and generate   finite number of physically
favoured
length scales.  These planewaves form finite
fields for prime factors $n$  of $p-1$.

\vm

There are also additional  arguments supporting the special  role
 of primes
near prime powers of two.\\
 a) If 2-adicity is
fundamental then 2-adic thermodynamics implies temperature quantization in
the form
$exp(1/T) \rightarrow 2^n$. An interesting possibility is that  the allowed
 2-adic
temperatures are not only integers but also primes: $n=p_1$. If  for  p-adic
primes $p$
Boltzmann factor is as near as possible to an allowed 2-adic Boltzmann
factor:  $exp(T)
\rightarrow p^n\sim 2^{p_1}$, one obtains     $p\sim 2^p_1$. \\
b) Fast Fourier transform is fastest, when the
  number of wavevectors is power of two so that spacetime volume has size
$2^mL_0$, where $L_0$ is discretization interval.

\subsection{Physical interpretation of $L_p$}

The tentative  interpretation of $L_p$  proposed in \cite{padTGD} was as a
lower cutoff length scale for p-adic  field theory.  The study of the low
 energy field
theory  counterpart of p-adic TGD \cite{lightfield} however
demonstrated that $L_p$ is in fact  infrared cutoff and UV divergences
cancel
without any momentum cutoff.  This results
give justification for the basic TGD:eish picture of  the topological
condensate.
 p-Adic
convergence cubes of size $L_p$ are the p-adic version of topological field
quanta introduced in \cite{TGD} and in certain sense largests 3-surfaces
 possible at a
given p-adic condensation level.  The topological condensation at level
$p_1>p$ is
possible and at this level these structures become particles perhaps
 describable using
field theory approach below length scales $L_{p_1}$.  Therefore the original
 assumption
is correct only in the sense that $L_p$ is a  natural  lower cutoff length
scale at
level $L_{p_1}$, when extended objects are treated as point particles.
One must
also  consider the possibility that the formation of larger structures by
gluing
p-adic convergence cubes together by tubes connecting together neighbouring
boundaries.   This mechanism is in
fact in key role in the  suggested  hadron physics,  nuclear physics and
 condensed
matter applications of p-adic TGD (see the third part of the book).  For
instance,
the color flux tubes connecting valence quarks can be identified as join
 along
boundaries bonds.

\vm

There is also a second, purely kinematical,  length scale generation
mechanism:  the construction of p-adic planewaves \cite{lightfield}
demonstrated that there is a preferred set of p-adic wavelengths, which are
proportional  to the  factors of $p-1$, whose number is relatively small for
 the
physically most  interesting primes $p$.

\vm

Length scale hypothesis leads to a picture of the  topological condensate as
 more
or less hierarchical structure consisting of almost flat 3-surfaces
$...<p_1<p_2...$ obeying effective p-adic topology so that the size of the
surface increases with $p$ and has rough order of magnitude given by $L_p$.
This concept of 3-space  replaces  ordinary 3-space with many sheeted
structure.   For instance, ordinary
macroscopic bodies can be regarded as 3-surfaces, whose outer boundaries
correspond to the surface of the body and boundaries are possible even in
 the
astrophysical length scales.

\vm

An obvious question is what happens above the length scale $L_p$. \\
a)  p-Adic
field theory description certainly fails.  The application
of  the length scale hypothesis in condensed matter length scales suggests
 that upper
bound for the  size of the  3-surface at condensate level $p$  is  typically
 of
order $ 10^2L_p$.  For nuclei and hadrons the length scale is of order few
 $L_p$.  \\
b) The basic condensation level for atoms is $p \simeq 2^k$, $k=131$ and
$L_{131}$ is
about one half of Bohr radius of hydrogen atom. The p-adic counterpart of
the hydrogen atom
seems to differ in many respects from real hydrogen atom. Coulomb  potential
 contains
factor $1/2\pi$ coming from the Fourier transform of photon propagator and
is not
p-adically well defined since $e^2$ is certainly rational so that  one must
 work in
momentum space and use discrete p-adic planewaves.  Because of $\pi$ the
 standard
expression for the energy levels of hydrogen atom does not make sense
p-adically.  The
p-adic versions of the  spherical harmonics do  exist (as homogenous
 polynomials of
coordinate variables)   and states with definite $J$ provide a good ansatz
 for
Schr\"odinger equation. There are number theoretic  problems with the
 definition of
energy $E=M-\Delta E$. The point is that rest mass defined as the  square
 root of $M^2$
thermal expectation value for electron at $k=137$ level  need not exist:
the reduction
of p-adic Schr\"odinger or  Dirac equation to the  standard form however
requires
this.   It is unclear whether the energies are sufficiently close to the
energies of
the  real hydrogen atom. All this suggests that real Schr\"odinger equation
 provides the
correct description of hydrogen atom. \\ d) Quite generally the success of
the  real
physics in long length scales suggests that  p-adicity is not  important in
length
scales above $L_p$. The generation of lattice structures in condensed matter
physics might be regarded as a direct manifestation of underlying
 p-adicity (basic
structural  unit being p-adic converge cube of size $L_p$) although real
physics
gives a good description of the situations.

\vm

All this suggests  that
although the p-adic condensation levels obey p-adic topology also in  length
scales $L>L_p$  real physics  must provide an excellent
description of physics in length scales $L>L_p$ and that p-adicity becomes
important
only below
 $L<L_p$.
The  only manner to avoid contradiction  is that ordinary  real physics
 works,  when
the length scale resolution is $L_p$, which in numerical practice means
discretized physics using rational numbers. In the more formal level this
means that
the effective action asscociated with action obtained by functionally
integrating in K\"ahler action degrees of freedom  with momenta larger than
$1/L_p$ defines in a good
approximation  spacetime with real topology. This hypothesis is in
 accordance with
the fact that real continuity implies p-adic continuity.
 The conclusion is in conflict
with  the original idea that length scale cutoff leads to effective p-adic
topology and more in concordance with the more standard assumption of
 p-adic physics
that p-adicity becomes important below some cutoff length scale $l$
 usually assumed
to be equal to Planck length scale.  Another manner to say the same
 thing is  that
quantum fluctuations in spacetime geometry are negligible in length scales
 larger than
$L_p$ so that  effective action obtained by averaging K\"ahler Dirac action
and
integrating over  small length scale degrees of freedom is real rather than
 p-adic.

\vm

An attractive possibility is that p-adic field theory  description at length
 scales
$L>L_p$ can be  based on discrete QFT  in the lattice formed by p-adic
convergence cubes and that the vertices are given by n-point functions of
 the
convergence cube.

\subsection{Physical interpretation of the  length scale $L_0$}

The fundamental p-adic mass  scale $L_0$  can be deduced from the
 experimental value of electron mass in p-adic description of Higgs
mechanism.

\begin{eqnarray}
L_0&=& \frac{\pi}{m_0}\nonumber\\
m_e&=& \frac{\sqrt{(5+\frac{2}{3}+\Delta )}m_0 }{\sqrt{M_{127}}}\nonumber\\
\Delta&\simeq &.0156
\end{eqnarray}

\noindent $\Delta$ corresponds to a small correction to the electron mass
 coming from Coulomb energy and topological mixing of lepton families.
 This gives

\begin{eqnarray}
L_0&=& \frac{\pi\sqrt{ (5+\frac{2}{3}+\Delta)} }{m_e   \sqrt{M_{127}}}
\end{eqnarray}

\noindent Using $ Gm_e^2\simeq  17.5145 \cdot 10^{-46}$ this gives

\begin{eqnarray}
L_0(\Delta=2^{-6}))&\simeq&1.824 \cdot 10^4 \sqrt{G}
\end{eqnarray}

\noindent  $L_0$  coefficient is by a factor $1.096$ larger than  its value
$L_0= 2^{10} k\sqrt{G}$, $k\simeq 16.25$   deduced in \cite{padTGD} from the
 first attempts to understand hadronic mass spectrum using p-adic super
conformal invariance.  Convenient reference scales are
 $L_{127}\simeq 2.89E-12 \ m$ and $L_{107}\simeq 2.82E-15 \ m$. The
definition of the length scales should
not be taken too literally since the definition necessarily contains some
unknown
numerical factor.  Rather remarkably,   the length scale $L_0$ is of same
order of
magnitude as the length scale at which the renormalized standard model
coupling
constants become equal and some kind of New Physics probably emerges.

\vm

The original hypothesis was that that the length scale at which p-adicity
 sets
 on is of order $L_0 \sim 10^4 \sqrt{G}$.   Above this length scale
spacetime should begin to look approximately flat piece of Minkowski space
 and
the field theoretic description should make sense.  This hypothesis  is
 probably not
 correct as the following arguments demonstrate.

\vm

  The basic problem is to understand the appearence of  the square root of
 $p$   instead of the naively expected $p$ in the expression of length scale.
 The first idea to come into mind is  that $M^4$ coordinate $m^k$ is
essentially the square root of p-adic spacetime coordinate $x^k$. The
 argument
was based on analogy with Brownian motion. $M^4$ coordinates as functions of
p-adic spacetime coordinates are analogous to the spatial  coordinate $x$  of
particle in Brownian motion and therefore one has

\begin{eqnarray}
\langle x^2(t)\rangle &=& Dt
\end{eqnarray}

\noindent  It has however turned difficult to realize this idea in
 Lorentz invariant manner.

\vm

An elegant   Lorentz invariant manner to understand the appearence of square
root is to assume that the p-adic version of  $M^4$ metric contains $p$ as a
scaling factor.

\begin{eqnarray}
ds^2&=& p m_{kl}L_0^2dm^kdm^l\nonumber\\
L_0&=&1
\end{eqnarray}

\noindent where $m_{kl}$ is the standard $M^4$ metric $(1,-1,-1,-1)$ and
$L_0=1$  holds true in units used and actually corresponds to  the distance
$L_0\sim 10^4 \sqrt{G}$.  The p-adic distance along k:th coordinate axis from
origin to  the point $m^k=(p-1)(1+p+p^2+...)=-1$ on the boundary of the
fundamental convergence cube of the square root function, is

\begin{eqnarray}
\sqrt{p}((p-1)(1+p+...))L_0&\rightarrow &\frac{(p-1)(1+p^{-1}+p^{-2}+...)}
{\sqrt{p}}L_0 = \sqrt{p}L_0=L_p\nonumber\\
\
\end{eqnarray}

\noindent  What is remarkable that     the shortest distance in the
range  $m^k =  1,..m-1$ is actually $L_0/\sqrt{p}$ rather than $L_0$ so that
p-adic numbers in range span the entire $R_+$ at the limit $p\rightarrow
\infty$.  What is remarkable is p-adic topology indeed approaches real
topology
in the limit $p\rightarrow \infty$ since the length of the discretization
 step
approaches to zero.  This intepretation is in accordance with the
 interpretation
of p-adic space time as effective spacetime surfaces spanned by most
probable
3-surfaces and with the intuitive idea that in very long length scales
 quantum
effects become vanishinly small so that effective spacetime surface becomes
continuous in real sense.

\vm

In TGD the 'radius'  $R$ of $CP_2$ is fundamental length scale and is of
 same
order as Planck length \cite{TGD}.  Concerning the formulation of conformal
 field
theory limit the basic problem is \\ i) whether one should can take $R$ as
basic
length scale and predict $L_0$ or \\ ii) whether  the value of $L_0$ is a
prediction of  Quantum TGD and  must be taken as given  in the
formulation of field theory limit.\\ The general physical picture is
in accordance with the latter alternative. This means that  in the
 expression of
the induced metric the parameter $R$ must be expressed in terms of $L_0$
and one
has following kind of expression

\begin{eqnarray}
R^2&=& kpL_0^2\nonumber\\
kp&\rightarrow &x\sim 10^{-6}
\end{eqnarray}

\noindent  $k$ is very large number, of order $10^{-6}p$.   It is
 illustrative
 to look how large the contribution of $CP_2$ metric to induced metric is.
  \\
i)    If the gradients of $CP_2$ coordinates are of order $O(\sqrt{p})$,
 the
 contribution of $CP_2$ metric to the  components of the  induced metric is
 of order
$kp^2$,  where as the contribution of $M^4$ coordinates is of order $p$.
 This
implies that the contribution if $CP_2$ metric to the distance is roughly
a
fraction  $x\sim 10^{-6}$ from the contribution of $M^4$ coordinates so that
gravitational effects  should be important immediately below elementary
 particle length
scales $L_p$.  This is not in accordance with the expectation that
 gravitation
becomes important in Planck length scales, only. \\ ii)  If the gradients of
$CP_2$  coordinates are of order $O(p)$ then the contribution of $CP_2$
 metric
to  the induced metric is of order $O(p)$ p-adically and in accordance with
 the
weakness  and even  the size of  the  coupling strength of gravitational
interaction for physically interesting primes $p$.

\section{Topological condensate as  minimum of effective action}

The construction of p-adic TGD necessitates  a satisfactory answer to the
question
'Why the  effective spacetime topology is p-adic?'.  One can consider  two
different answers but only one of them seems to be entirely satisfactory
 and is
 based
on the observation that observed spacetime  is actually some kind of quantum
average defined as minimum of the  effective action.

\subsection{p-Adicity from length scale cutoff?}

Consider first the wrong answer.
 The original explanation for the p-adicity of  effective spacetime topology
 was
 in terms of length scale cutoff.
The concept of the  length scale cutoff plays fundamental
role in the
description of the critical phenomena. The TGD: eish counterpart of
this procedure is the smoothing out of the topological details of
size smaller than some given length scale $L$.  This procedure  should
give the 3-space of General Relativity: the  presence  of details is
described by energy momentum tensor and
various particle densities.  The difficulty is that the usual momentum
cutoff of quantum field theories doesn't  provide a solution to this
problem.  The geometric description of this procedure provided by
various lattice models seems to be more in the spirit of TGD:eish
description of the  cutoff procedure.

\vm

Smoothing out procedure  (or coarse graining) need not respect
differentiability for
the simple reason that differentiability in Planck length scales
 does not correlate the
values of functions at distances much larger than Planck length.
This suggests that
smoothed out spacetime surfaces obeys p-adic topology instead of real
topology.    A
possible definition for  the  smoothing out  procedure is as p-adic
cutoff:
replace $x_p$ appearing as the argument of $h^k$   with its first $N$ pinary
 digits.
This means that all details of the  3-surface related to the higher
pinary digits
of the expansion are smoothed out: by p-adic continuity these details
are the smaller the higher the corresponding pinary digit is.
The presence of the details must be described somehow and the use of p-
adically differentiable densities is a natural manner to do this.

\vm

This interpretation need not be  the correct one. \\
a)  The appearence of the  favoured p-adic length
scales
corresponding to powers of $p$ and also the appearence of  the favoured
 values of
$p$ is not in accordance with  the arbitrariness of p-adic cutoff length
 scale.\\
b) p-Adic QFT turns out to be  UV finite without any p-adic length scale
 cutoff.\\
c) Despite several attempts it has not possible to define precisely what
the concept
RG transformation and RG invariance of K\"ahler action means in this
 interpretation.

\vm

In fact, in light of the interpretation to be  suggested below the smoothing
 out
procedure
 should lead from p-adic topology to real topology, when the  cutoff length
scale is  larger than $L_p$ rather than from real to p-adic!

\subsection{p-Adic spacetime as absolute minimum of effective action}

In TGD the geometry of 3-surface and the corresponding unique classical
 spacetime
surface are subject to quantum fluctuations and  the criticality of  the
 TGD:eish
Universe suggests that these  fluctuations
are important in all length scales.   Therefore the  observed 3-space and
spacetime should
correspond to nontrivial quantum averages.  These averages obviously
depend on quantum
state.
 p-Adic topology could  be one of the  basic  properties of this average
spacetime.

\vm

The problem is to formulate the concept of quantum average for 3-space and
 spacetime.
In ordinary QFT the concept of effective action is in decisive role.
Effective
action is typically obtained by coupling some fields appearing in the
action
 exponential
to external currents and by  functionally integrating over some fields
 appearing in
the action exponential to get effective action in terms of new field $\Phi$.

\begin{eqnarray}
\Phi_{new}&=& \frac{1}{\Gamma}\frac{\delta  \Gamma}{ \delta J(x)}\nonumber\\
 \Gamma&=& \int exp(S(\Phi)+ J\Phi) D\Phi \equiv exp(S_{eff}(\Phi_{new}))
\end{eqnarray}.

\noindent The absolute minima of the effective action define the vacuum
average of the field $\Phi (x)$ in presence of external current $J$.

\vm

In TGD the corresponding construction could proceed as follows.  K\"ahler
 Dirac
 action
defines configuration space geometry.  What one is interested in  is the
 integration
over
configuration space degrees of freedom to get effective action, whose
 absolute
 minima
would define p-adic space time concept.  The effective spacetime is
characterized
by $H$ coordinates $h^k_{new}(x)$ and the simplest coupling of these
 coordinates
 to ordinary
H-coordinates is obtained by coupling K\"ahler potential to the external
 K\"ahler $j^{\alpha}_{new}$
current defined by $h^{k}_{new}$

\begin{eqnarray}
S_K(h)+S_D &\rightarrow& S_K(h) +S_D + \int  j^{\alpha}_{new}(x)
 A_{\alpha}(x)
\sqrt{g}d^4x\nonumber\\
j^{\alpha}_{new}&=& (D_{\beta}J^{\alpha\beta})_{new}\nonumber\\
J^{\alpha\beta}_{new}&=& g^{\alpha\mu}_{new}g^{\beta\nu}_{new}
J_{\mu\nu}^{new}
\end{eqnarray}

\noindent Here the subscript 'new' means that the induced K\"ahler field and
induced metric
are determined entirely by the new coordinates $h^k_{new}$. The matrix
element
of the exponential of the effective action in a given quantum state is
 defined as
the average
obtained by integrating over configurations space degrees of freedom that is
over
all  possible 3-surfaces.  It must be emphasized that the resulting effective
action depends on spinor fields since one doesn't integrate over them.
Effective spacetime is determined as the absolute minimum of effective
 action
and need not  be unique in a given quantum state:  in thermodynamics the
nonuniqueness means possibility   of several phases  and criticality
suggest the existence of a large degeneracy.

\vm

A comment on the treatment of boundary components in quantum averaging
is in order.  The success of the  ordinary QFT suggests that boundary
components can be replaced with quantum averages in vibrational degrees of
freedom but averaging does not make sense in cm degrees
of freedom.  Quantum  averaging should give rise to some boundary 3-surface
 or a family
of 3-surfaces (degeneracy is possible) not depending much on the details of
 the
surrounding spacetime. Degeneracy is indeed expected and corresponds to
conformal and modular degeneracy.  The idealization of the boundary
 component with a
point like particle looks rather natural so that effective spacetime
contains
boundaries as wordlines.  Kac Moody spinors
provide economical manner to describe conformal degrees of freedom.  The
coupling of
the  wordline dynamics to interior dynamics is  in principle determined by
 the boundary
conditions on the boundaries and the  conservation of various charges.

 \vm

What  can  one conclude about the general form of effective action?\\
a) RG invariance of the effective action has been one of the basic
 hypothesis  of
Quantum TGD \cite{TGD}. The problem is to define this concept precisely.
 For
simplest RG invariant actions correspond to free field theories and for
 these
theories effective action is identical with the original action. This
observation suggests a  precise definition for RG invariance:  the  effective
action associated with K\"ahler action in the absence of spinor fields is
 indeed
formally identical with the K\"ahler action but the effective spacetime
topology need
not be the real one.  The same conclusion is achieved if one assumes that
 effective
action possesses same approximate symmetries and the characteristic vacuum
degeneracy
selecting K\"ahler action uniquely.
 \\ b)  An attractive possibility is that the  effective
spacetime   topology is that of a   hierarchical topological condensate
$...<p_1<p_2...$ consisting of pieces possessing p-adic topology with
various
values of $p$. The gluing of two topologically different pieces $p_1$
and $p_2$
should be  performed along their boundaries. This gluing
operation can be performed using canonical correspondence.   \\ c) The
enormous vacuum
degeneracy of the  K\"ahler action and associated exact $Diff(M^4)$ and
canonical
invariances suggests analogy with spin glass phase in the sense that there
exists enormously large number of vacuum spacetimes serving as background
 for
fermionic dynamics. The averaging over these vacuum spacetimes is
 completely
analogous to the averaging over  the bonds strengths in spin glass phase.
 The basic feature of the  configuration space of minima in   spin glass
 phase is
ultrametricy. Ultrametricity is  the basic feature of p-adic topology and
therefore the
 topological condensate as a   hierarchy of p-adic topologies $p_1<p_2<...$
is
perhaps the most general realization of the ultrametricity. \\
d) What about the form of effective action, when spinor fields are
nonvanishing?
RG invariance suggests that the general form remains invariant but something
happens for $H$-coordinates appearing as arguments of the action. TGD allows
$N=1$ supersymmetry generated by covariantly constant right handed neutrino
and this suggests the solution to the problem. What happens is that
H-coordinates are replaced by super coordinates
$h^k \rightarrow h^k+ K \bar \Psi
\gamma^k \Psi$.  The replacement is coordinate invariant  only provided one
 can select
preferred coordinates for $H$. For  $M^4$ the standard Minkowski coordinates
are obviously the preferred coordinates. For $CP_2$ the complex coordinates
 for which
the action of $su(2)\times u(1)$ is linear   are preferred coordinates.
These coordinates are defined only modulo color  rotation and consistency
requires that physical states are color singlets. This implies that each
p-adic region
each color singlet so that color confinement in strong sense  results.

\vm

The interpretation of the  p-adic spacetime as
effective quantum spacetime is in principle possible in arbitrarily short
 length
 scales
and for all spacetime topologies and the approximate symmetries   are not
 essential
for this in this respect. Later
it will be found that p-adic conformal and canonical invariances seem to be
 completely
general approximate symmetries of the zero action extremals of the effective
 action
broken by gravitational effects only.  Therefore p-adic spacetime concept
 is not
just a convenient approximation working at the flat spacetime limit as
 originally
believed    but an  essential part of
Quantum TGD.  The only approximation made in the  p-adic field  theory limit
 is the
freezing of the  spacetime degrees of freedom by using quantum averaged
spacetime
determined by the effective action and it might be even possible to develop
 a formalism
to calculate the corrections to this approximation.

\subsection{The value of the  K\"ahler coupling strength?}

The requirement that the  effective action is identical with  the  super
symmetrized
K\"ahler Dirac action should fix the value  of K\"ahler coupling strength
$g_K^2$ and thus make theory unique.  There are two independent heuristic
arguments leading to
same estimate for the K\"ahler coupling constant. \\
a) Coupling constant evolution argument \cite{TGD} led to the estimate

\begin{eqnarray}
\alpha_K&=& \frac{28}{27}\alpha_{em}(low) \nonumber\\
\alpha_{em}(low)&\simeq& \frac{1}{137}
\end{eqnarray}

\noindent b)
 The experience with  the ordinary field
theories suggests that elementary particle mass  scale is related to the
exponent of
the inverse of $\alpha_K$: $L_{elem}=exp(c/\alpha_K)/\sqrt{G}$, where $c$ is
 some
constant. A generalization of this formula is suggested by two
observations: \\
i)
Mersenne prime $M_{127}$ defines the  largest elementary particle length
scale and is also
a member in the Combinatorial Hierarchy $M(n)=M_{M(n-1)}$ of the  Mersenne
numbers
(possibly all primes)  given by
$3,7,127, M_{127},..$.  \\
ii)  The exponent of the  K\"ahler action for $CP_2$
type extremals describing elementary particles defines natural large
 dimensionless parameter of order $10^{19}$  and in \cite{TGD}  an
 explanation
 of  the mass elementary particle mass scale in
terms of this number was suggested.\\
These observations suggest the following formula for the
fundamental length scale $L(127)$  as a  prediction of TGD:

\begin{eqnarray}
L(127)&= &exp(S_K(CP_2))\sqrt{G}\nonumber\\
L(127)&=& \sqrt{M_{127}} L_0\nonumber\\
S_K(CP(2))&=& \frac{\pi}{8\alpha_K}\nonumber\\
L_0 &=& k 10^4\sqrt{G}\nonumber\\
k&\simeq& 1.824
\end{eqnarray}

\noindent These formulas give

\begin{eqnarray}
\alpha_K &=&.9641 \frac{137.03604}{\alpha_{em}(low)}  \cdot
\frac{28}{27}\alpha_{em}(low)
\end{eqnarray}

\noindent The two estimates give identical results  if one has
$\alpha_{em}(low)= .9641/137.03604$: this assumption is  consistent with
the
electromagnetic coupling constant evolution and gives $\alpha_K=137.06$.
$\frac{1}{\alpha_K}=137$
   would give $k=1.77<1.824$, which is
however too small value.
 In any case,  the two
heuristic arguments  are consistent which each other and internal
consistency predicts the sale $L_0$ correctly.  $\alpha_K$ has a
 natural upper  bound
corresponding to $L_0=\sqrt{G}$ and given by $\alpha_K\geq .0089$. The
great task is to
understand the generation of the length scale $L_0$ or equivalently the
low energy
value of the fine structure constant.

\vm

It is quite possible that conformal
invariance gives conditions on the value of $g_K$.
 The construction of  unitary representations of conformal symmetry  should
 lead to a definite value for K\"ahler coupling $g_K$ (conformal central
charge
$c$ vanishes).  The  conformal charges contain parts coming from
Dirac action and from K\"ahler action and the part coming from K\"ahler
action
depends is proportional to  $1/g_K^2$.  Therefore the requirement that
 conformal
current commutators contain no anomalies and that vacuum expectation of
 Virasoro
generator $L_0$ should have quantized value fixed by unitarity,  could  pose
 condition
on the value of  K\"ahler coupling strength.

\subsection{Approximate  symmetries of the  p-adized
K\"ahler action}

The criticality together with two-dimensional analogy suggests that
4-dimensional conformal transformations act as approximate symmetries
of p-adic   K\"ahler action. Also the existence of p-adically analytic
extrema
of p-adic K\"ahler action suggests the same.    Actually the symmetry
group is much larger.  When gravitational effects can be neglected
 $Diff(M^4) $
acting on $M^4$ coordinates is approximate symmetry. Besides
this  canonical transformations of $CP_2$ localized with respect to $M^4$
coordinates act as dynamical symmetries in the interior of the spacetime
surface. The emergence of $Diff(M^4)(M^4)$ suggests that the theory does not
 differ
much from what might be called topological field theory in the interior of
spacetime surface. It however turns out that p-adic quantization of Dirac
 action
allows massless states, presumably including also graviton. Actually these
 states
correspond to the  representations of certain Super Virasoro algebra with
the property
that all generators except $L_0$  annihilate vacuum whereas $L_0$ gives the
energy of the (massless) state.

\vm

 p-Adic diffeomorphisms  acting
 on $M^4$   via the formula
$M \rightarrow W(M)$ and canonical transformations of  $CP_2$ local with
respect to $M^4$  act as
 exact
symmetries of
 p-adic K\"ahler action,  when
gravitational effects are neglected. This motivates the realization  of these
symmetries as exact symmetries by requiring that conformal and canonical
charges annihilate physical states.

\vm

   The change of the action in general
transformation can be decomposed into two parts:
 $\delta S= \delta S_1+\delta
S_2$ corresponding to  the change induced by the change of  the
 induced K\"ahler
form and of the metric.   \\
  a) For the extremals of K\"ahler action  any variation of the action
 reduces to
total divergence, which is  the divergence of the current
associated with infinitesimal transformation and the vanishing of this
divergence  means just charge conservation.   In present case the total
divergence contains only the contribution
 $\delta S_1= T^{\alpha\beta}\delta
g_{\alpha\beta}$  coming from the change of the  induced metric since
the induced
K\"ahler form remains invariant for both diffeomorphisms and canonical
transformations.  The contribution coming from the  metric is in general
  very
small  small by the weakness of
the gravitational  interaction and vanishes in the approximation that
 $CP_2$
contribution to the induced metric is zero. \\
 b)  The idea is to    pose the vanishing  of the
total divergence as a quantum
 condition
on the physical states: this condition states that $Diff(M^4)$ (or perhaps
 only
conformal)  and canonical (local with respect to $M^4$)
 charges are
conserved. A more stronger condition  that these charges
 annihilate physical states  is  indeed the basic constraint
 on physical states
in  field theories. The condition makes sense as a condition for
the  effective action since the minima of the effective action  can be
 regarded
 as effective spacetimes associated with quantum states. Also the breaking
 of
conformal
symmetry in the sense that $L_0$ is conserved but  does not annihilate
 physical
states is  possible.

\vm

The  symmetries defined in question  are
extremely general symmetries as the following examples demonstrate. \\
a) $Diff(M^4)$ and local canonical symmetries are exact symmetries in the
usual
sense  for the
 vacuum extremals: any
4-surface with $CP_2$ projection, which is Lagrange manifold
 (having vanishing induced K\"ahler form). Lagrange manifolds are
in general 2-dimensional, which means that  a very rich variery of
topologies
becomes possible for the vacuums. This vacuum degeneracy implies the analogy
between TGD and spin glass.  Diffeomorphisms of $M^4$ are
symmetries in the vacuum sector.\\ b) Spacetimes representable as maps
$M^4\rightarrow CP_2$ and carrying  nonvanishing K\"ahler field and
 satisfying
Maxwell equations $j^{\alpha}=0$ are
 examples
of nonvacuum extremals. An interesting possibility is that also the
 p-adically
 analytic
maps $M^4\rightarrow CP_2$ to be described in next section define actually
exact
solutions of the  effective action. One example of zero action minima of
 K\"ahler
action are so called massless solutions  discussed in \cite{TGD}. These are
defined as maps $M^4\rightarrow CP_2$

\begin{eqnarray}
s^k &=& f^k(p\cdot m, \epsilon \cdot m)
\end{eqnarray}

\noindent where $p$ is massless four-momentum vector $p\cdot p=0$ and
 $\epsilon$
is polarization vector orthogonal to $p$: $\epsilon \cdot p=0$ and the
functions
$f^k$Ê are arbitrary. In this case canonical transformations are exact
symmetries. \\ c) String like objects of type
$X^2 \times Y^2\subset M^4\times
CP_2$ are vacuum solutions and suitable deformations of these objects with
vanishing action and nontrivial K\"ahler form probably define nonvacuum
extremals with vanishing action.   The identification of hadronic strings as
objects  of this type was suggested in \cite{TGD}.  The hadronic string
tension
would be generated dynamically and be due to the presence of fermions and
nonvanishing K\"ahler fields. In fact, classical gluon field is proportional
to
K\"ahler field: $G^A \propto H^AJ_{\alpha\beta}$, where $H^A$ is the
Hamiltonian of color   transformation.  It should be noticed that color
symmetries are canonical transformation so that physical states must be
 color
singlets also  by the  canonical invariance. \\ d) The so called $CP_2$
type extremals
obtained by deforming the standard imbedding of $CP_2$ so that the $M^4$
projection of the surface is light like curve

\begin{eqnarray}
m^k& =&f^k(s)\nonumber\\
m_{kl}\frac{df^k}{ds}\frac{df^l}{ds} &=&0
\end{eqnarray}

\noindent These surfaces are metrically equivalent with $CP_2$ and are
 vacuum
solutions. Canonical and conformal symmetries are exact symmetries for
these
surfaces.  In \cite{TGD} it was suggested that elementary particles
correspond to
$CP_2$ extremals and that Feynmann diagrams correspond to surfaces obtained
 by
gluing these extremals together by connected sum operation. The action of
 these
solutions is nonvanishing and these surfaces as such are  not absolute
 minima of
effective action.  One can however consider more general solutions obtained
as a
topological sum of $CP_2$ type extremal and background spacetime surface
carrying nontrivial K\"ahler Coulomb field, which gives a negative
contribution
to action so that total action and its real counterpart  vanishes and
 one obtains
absolute minimum. The exponent of K\"ahler action defines in natural
manner elementary
particle length scale as suggested already in \cite{TGD}.

\vm

These examples encourage the conjecture that the field theory limit of TGD
defined as a
fermionic field theory in  the average spacetime defined by the physical
state  allows local
canonical transformations  and $Diff(M^4)$   as
gauge symmetries under rather general conditions. The approximate
invariances are
 not  needed in the  definition of the  super symmetrized K\"ahler action.
 The
surprising success of the mass calculations based on super conformal
invariance also
supports the  realization as gauge symmetry. It might well be that these
symmetries are
spoiled 'in large' by gravitational effects. A purely p-adic feature of
these
symmetries is  however that one can restrict the symmetries to be
 sufficiently small to
force the change in the induced metric small enough. This is achieved if
 the
parameter $t$ in the exponential  $exp(tJ)$ of Lie-algebra generator has
 p-adic norm
smaller than $p^{n/2}$: these tranformations indeed form a group for any
integer $n$.
In this manner the configuration space decomposes into disjoint blocks and
 one obtains
different realization of symmetries inside each block. This is all what is
 need to
deduce the most important consequences of conformal invariance (such as
univesality
of  the  mass spectrum). These disjoint blocks are clearly analogous to
 the degenerate
vacua of the  spin glass phase.

\subsection{ Gravitational warping}

The absolute minimum of effective action depends on the quantum numbers of
the
state.  In particular, there is dependence on four-momenta and various
charges of the quantum state.   Gravitational warping represents perhaps the
simplest dependence on the quantum numbers of the state.
 Gravitational warping is completely
analogous  with   the existence of infinitely many imbeddings of flat
2-dimensional space  $E^2$ to $E^3$ obtained by deforming any plane of
 $E^3$
without stretching it.  The  physical reason for the  warping is the
gravitational
mass of the p-adic convergence cube.   The mass of the p-adic convergence
cube deforms spacetime surface and average effect is  momentum dependent
scaling of some metric components.
   In GRT gravitational warping clearly does not make any
sense. The simplest  physical consequence of the deformation is that photons
propagating in condensate spend in general longer time than vapour phase
 photons
to travel a given distance \cite{TGD}.

\vm

  A rather general nonstandard imbedding of $M^4$ representing gravitational
warping is obtained by assuming that  $CP_2$ projection of spacetime surface
 is
one-dimensional curve:

\begin{eqnarray}
s&=& KpP_km^k \nonumber\\
g_{\alpha\beta}&=&  m_{\alpha\beta}-K^2p^2 P_{\alpha}P_{\beta}\nonumber\\
m_{\alpha\beta}&=& p \eta_{\alpha\beta}
\end{eqnarray}

\noindent  Here $s$ is the curve length coordinate,   $P_k$  is total
four-momentum and one has  $\vert K\vert =1$.

\vm

The  scaling  of the metric components has two nice consequences as far as
the convergence of the perturbation theory is considered:\\ a) Assuming that
the spectrum of off-mass shell four-momenta for spinors is  not changed the
propagator expression has no poles since one has always
  $k^2= g^{\alpha\beta}k_{\alpha}k_{\beta}\neq M^2$
in warped metric. \\
b)    For
$m_{00}=1\rightarrow  1-\delta$  the    expression $k_{\perp}^2=0$  becomes
$k_{\perp}^2=-\delta k_0^2$ and this number becomes p-adically very large,
 when
p-adic norm of $k_0$ increases. This guarantees the convergence of the
propagator expression as well as finite degeneracies for all values of $P$.

\vm

Gravitational warping cannot make spacetime metric  Euclidian if boundary
contribution to the mass squared is given by  the standard stringy mass
formula. In the
rest frame the deformation of the metric is just $ g_{00}= p(1-K^2pM^2)$.
Stringy
mass spectrum is of form $M^2=kn$, $\vert k\vert_p =1$ so that the  p-adic
norm of
the mass squared is not larger than one and the contribution  to  the metric
 is of order
$O(p^2)$ p-adically.  It is worth noticing that stringy mass formula gives
upper
bound for the mass contained in a region of size $L_p$: $M\leq M_{max}\sim
\sqrt{p}10^{-3}M_{Pl}$. An interesting possibility is that stringy mass
formula
might hold true even in astrophysical length scales.  For Sun $L_p$ is of
order
of the size of solar system and upper bound for mass is much larger than
 solar
mass. For neutron star of size of order $10^4$ meters the mass is not far
from
its  maximum value for  $L_p\sim 10^4 m$.

\subsection{ Idealization of   boundary components with point like objects}

In TGD boundary components of 3-surface are carriers of elementary particle
 quantum numbers.
In particular,  the state of elementary particle  depends on modular degrees
 of freedom of the boundary component. In \cite{TGD} the  so called
elementary
particle vacuum  functionals were constructed and the p-adic description of
Higgs mechanism relies   on the p-adized version of these functionals.
A purely
p-adic feature  of the construction is  the discretization of p-adic moduli
space.  The route from ordinary modular degrees of freedom of Quantum TGD
to
discrete p-adic modular degrees of freedom  of p-adic TGD should take place
 via
the construction of 'effective'   elementary particle functionals by
coupling
modular degrees of freedom to the p-adic modular degrees of freedom somehow.
How this process is carried out in detail is not considered here.

 \vm

As already found one cannot perform quantum averaging for the cm degrees
of
freedom of boundary component and p-adic spacetime surface is the average
 spacetime
obtained by keeping the center of mass positions of boundaries at some
reference time fixed.    The practical treatment of boundary degrees of
 freedom
 in  the low energy  field theory limit suggests the idealization of
boundary
component  with point like object.  This necessitates quantum mechanical
averaging  over   the modular degrees of freedom.   This quantum mechanical
averaging gives to the  mass squared the dominating modular contribution.
 The
modular contribution to mass squared is indeed  quantum mechanical  rather
 than
thermal average: this issue was left open in  the p-adic description of
 Higgs
mechanism.

\subsubsection{Induced Dirac spinors on the world line}

  The simplest
description of the  point like limit is obtained by neglecting totally the
vibrational degrees of freedom.  The  3-dimensional orbit of the
boundary component is replaced with the  world line of a  point like
 particle and  the
Dirac action for the induced spinors  on the  boundary component is replaced
 with the
Dirac action on  with world line.    The induced spinor connection   on
 the world line
reduces to a pure gauge with a suitable choice of $CP_2$ vielbein basis
and a
suitable gauge for  the K\"ahler potential on world line and one can wonder
 whether
the  couplings of the fermions  to the  classical gauge fields are actually
 trivial
(note that the emission of a gauge boson corresponds to the splitting of the
 boundary
component world line).  This is not probably not the case:  the point is that
boundary component cm degrees of freedom are not frozen and one must treat
 these
degrees of freedom quantum mechanically so that  there is no general gauge
choice
tranforming the gauge potentials away.
  If  the couplings of  the fermions to the
classical background fields were trivial  the problems
associated with the  classical long range $Z^0$ fields (the possibility
 large parity
breaking effects)  considered in the third part of the book would have
 an easy solution.
Physical intuition and the previous work with TGD  however suggest that
 classical
background fields are important.

\vm

 The contribution of the  modular degrees of freedom to mass squared
should be  taken into account somehow.  The only manner  seems to be as
 a constraint on
the effective mass squared of spinors.  The Dirac equation for induced
spinors contains
a term proportional to the trace of the second fundamental form of the
worldline

\begin{eqnarray}
\Gamma^{\mu}D_{\mu}\Psi &=& \frac{1}{2} \Gamma_k H^k\nonumber\\
H^k&=& g^{tt}D_tD_th^k
\end{eqnarray}

\noindent  $H^k$ decomposes into $M^4$ and $CP_2$ contributions and the
 description of Higgs mechanism proposed in \cite{TGD} was based on the
identification of $CP_2$ part of $H^k$ as Higgs field.   If $H^k$ is
covariantly
constant quantity along the world line.  It seems natural to require that
 the
square of $CP_2$ part $S^k$  of $H^k$ is equal to the modular contribution
 to
the mass squared on the world line

\begin{eqnarray}
M^2(mod) &=& S^kS_k
\end{eqnarray}

\noindent In general $S^k$ is not covariantly constant but one could perhaps
 add the covariant constancy of $S_k$ along world line as an additional
condition:

\begin{eqnarray}
D_tS^k&=&0
 \end{eqnarray}

\noindent  The use  of  the  constraints is in accordance with the concept
 of the
effective action. What constraints express is the state dependence of the
average spacetime surface on quantum numbers. An additional constraint comes
from the requirement that electromagnetic gauge group, which is subgroup of
local $so(4)\times u(1)$ is exact gauge group. This implies that induced
$W^{\pm}$ fields mixing different charge states of fermion vanish on world
line. Altogether this means that the $CP_2$ projection of  the world line
 is
 highly determined: a detailed consideration can be found in context of
the symmetries.  The  motion in $M^4$ degrees of freedom is determined by
 the
boundary conditions stating  the conservation of four-momenta and other
 charges.

\subsubsection{Generalization of Kac Moody Dirac operator to worldline}

The simplest point like limit implies that spinors have only finite number
 of
degrees of freedom and therefore a loss of conformal degrees of freedom
necessary to obtain Planck mass excitations.  According to the mass
calculations
elementary particle masses  contain besides the dominating  modular
contribution  a contribution from spin, electroweak and color degrees of
freedom, which can
 be calculated using p-adic thermodynamics with  energy replaced with the
Virasoro generator $L_0$.
 Therefore it seems to be an overidealization to
eliminate totally the conformal degrees of freedom. Of course, one could use
 the total mass squared including the thermal
contribution in the Dirac  equation for ordinary induced spinors.  Probably
this is not wise.
The mass calculations
provide a better solution of the problem: replace $H$-spinors with the
 Kac-Moody
spinors constructed as representations of Super Virasoro algebra
\cite{padmasses}.  One can define the coupling of
the  world
line gauge fields to Kac Moody spinors and also the coupling  of
 $H^k\Gamma_k$  is
well
defined and due to covariant constancy is combination of  the flat space
gamma matrices.
The possibility to eliminate neutral gauge potentials by gauge
 transformations
suggests  that in $CP_2$ degrees of freedom actually no dependence on
 $CP_2$
coordinates remains and KM spinors in $CP_2$ degrees of freedom behave as
 flat
space spinors in cm degrees of freedom.

\vm

p-Adic description of Higgs mechanism led to the concept of Kac Moody Dirac
operator acting on Ramond or N-S type representation. For Ramond type
representation
the operator is sum of three terms.

\begin{eqnarray}
D&=& p^k\gamma_k(M^4) + M_{op}+G_0\nonumber\\
M_{op}&=& b_k\gamma^k(CP_2) \nonumber\\
M^2_{op}&=&M^2+ M^2_{mod}
\end{eqnarray}

\noindent  The first term is   $G_0$ acting on Super Virasoro, presumably
 acting on conformal degrees of freedom
of two-dimensional boundary component defined as $m^0=constant$ slice of
 3-dimensional
boundary. The second term is Dirac operator in $M^4$ cm degrees of freedom.
The third term is mass operator, which is linear combination of $CP_2$ type
 gamma matrices
anticommuting with $G_0$ and is required to give correct conformal weight
for states, which should be massless in absence of the modular contribution
$M^2_{mod}$  to mass
squared
 (second paper of \cite{padmasses}). The presence of
$M_{op}$ can be
 interpreted
as a breaking of the  super conformal symmetry.

\vm

 What is nice is that the
tachyonic vacuum weight of $L_0$, which remained mysterious in the mass
calculations,
can  be understood in terms  of  the induced spinor concept.
 A second  great mystery of the mass calculations was   the origin of the
 additional cm
part in $G_0(tot)$, in particular mass operator. It  would be very nice
if $G_0(tot)$
would actually correspond to Super Virasoro
 generator.
This is indeed the case if Super Virasoro generators for the  boundary
component
idealized to
point are defined as sums of two terms. The  first term corresponds to the
 Super
conformal transformations
of proper time coordinate for which induced metric $g_{tt}$ is constant.
Second
 term corresponds
to the Super Virasoro generator acting on  the two-dimensional boundary
component
 in standard
manner.
If  proper time is used  the  super generators
$G_n$ $n>0$ annihilate the state automatically and $G_n(tot)$
 reduces to
$G_n$.  $G(tot,t)$ defined as

\begin{eqnarray}
G(tot,t) &=& \sum_n G_n(tot)a^{nt}\nonumber\\
G_n(tot,t)&=& G_n(worldline) + G_n(bound)\nonumber\\
G(worldline,t)&=&\frac{1 }{2}(\gamma^tD_t^{\rightarrow}-
D_t^{\leftarrow}\gamma^t)
\end{eqnarray}

\noindent  however reduces to $G(bound,s) $ in proper time frame $t=s$
 apart
from the cm part

\begin{eqnarray}
G_0(wordline) = p_s\gamma^s &=&  \gamma_k(M^4)\frac{dm^k}{ds}+
\gamma_k(CP_2)\frac{ds^k}{ds}+...  \nonumber\\
&\equiv& p_k \gamma^k(M^4) + p_k(CP_2)\gamma^k(CP_2)
\end{eqnarray}

\noindent If $t$ is  proper time the contribution is indeed constant.

\vm

The proper definition of the fermionic KM Dirac action density on the world
line
is therefore as

\begin{eqnarray}
\bar{\Psi} G(tot,t)\Psi &=&\bar{\Psi}(G(worldline,t)+ G(bound,t))\Psi
 \nonumber\\
&=&\bar{\Psi} (\gamma^s \frac{d}{ds}+ G(bound,s))\Psi
\end{eqnarray}

\noindent where the last expression holds true in proper time coordinate
 $s$.
KM Dirac equation states that

\begin{eqnarray}
G_0(tot)\Psi&=& -M_{op}\Psi\nonumber\\
M_{op}&=& \frac{H^k\gamma_k}{2}\gamma_k\Psi
\end{eqnarray}

\noindent  The mass term $M_{op}$ comes from the fact that
induced gamma matrices are not covariantly
constant: $D_{\alpha}\Gamma^{\alpha}=H^k\gamma_k$, where $H^k$ are the
 components of
the second fundamental form of the  world line.  The breaking of  the
 conformal symmetry
via the generation of a  nonvanishing negative (tachyonic) vacuum conformal
 weight
$h(vac)$  requires that $M_{op}$ is nonvanishing. The value of $h(vac)$
is $-2$ ($-1$)
for isospin $1/2$ ($-1/2$)  fermions and $-5/2 $ for bosons.  The
presence of this  term implies that world line cannot be a geodesic
line $CP_2$ degrees
of freedom. In $M^4$ degres of freedom accelerated motion causes additional
 breaking of
conformal symnetry and small change in mass squared.  If second fundamental
 form is
covariantly constant then only the mass of the particle  changes  and
higher  Virasoro
conditions remain true.
 Although electroweak doublet Higgs particle
is definitely   absent from the mass spectrum of the theory, second
 fundamental form
however appears in the KM  Dirac equation in the role of classical
 Higgs field so that
the original identification of the  Higgs field as the trace of the second
fundamental
form is not totally wrong.

\vm

The treatment of N-S case proceeds in similar manner. The  new feature is
that KM Dirac operator has conformal weight $1/2$ and corresponds to
$G^{1/2}(tot)$ for world line containing ordinary conformal part and the
part
acting on the world line:

\begin{eqnarray}
G(tot,t)&=& G(worldline,t) + G(bound,t)\nonumber\\
G(tot,t)&=&\sum_n G(tot,n+1/2)a^{(n+1/2)t}\nonumber\\
G(wordline) &=& \gamma^t D_t^{\rightarrow} - D_t^{\leftarrow}\gamma^t
\end{eqnarray}

\noindent  In bosonic case one expands $G(tot)$ with respect to half odd
integer
p-adic planewaves on world line.   Bosonic KM action density on world line
is of
form

\begin{eqnarray}
L_B&=&\bar{\Psi} G^2(tot,t)\Psi
\end{eqnarray}

\noindent If propertime coordinate is used  and second fundamental form is
covariantly constant equations of motion reduce to  Virasoro conditions for
$L_n$, $n>0$ and $n=0$ term gives breaking of conformal symmetry.

\vm

 Virasoro
conditions are automatically true if $G_n$,$n>1/2$  annihilates the states
automatically and this must be assumed in order to achieve Super Virasoro
invariance.
  Again the equation of motion
implies additional mass squared term with geometric interpretation

\begin{eqnarray}
G^{1/2}(tot)\Psi&=& -M^{1/2}_{op}\Psi\nonumber\\
M^{1/2}_{op}&=& -\frac{\gamma_k^{1/2}H^k}{2}
\end{eqnarray}

\noindent  The operator
$M^{1/2}_{op}$ is expressible as linear combination of $M^4$ gamma matrices
and
$CP_2$ type gamma matrices $\gamma^{1/2,k}$, $k=0,3$ and the square of
$M^2_{op}$ is such that operators with conformal weight $h=5/2$ create light
states (mass comes only from modular contribution to the mass of higher
genus
bosons).

 \subsection{Second quantization of the  field theory limit
and p-adic thermodynamics}

p-Adic  field theory limit is not all what is needed to construct a
realistic theory.  One must also perform  second
quantization both in interior and boundary degrees of freedom. Although not
very
much will be done in this direction in this paper it is useful to have rough
 idea
about what is involved.

\vm

Consider first second quantization in boundary degrees of freedom.\\
a)  The basic point is that  field theory inside p-adic
convergence cube  allows only very few light states for both the interior
and
boundary components of   single convergence cube and in practice the light
excitations of  boundary components must correspond to elementary particles.
 \\
b)   In practice single boundary component can in a good approximation be
idealized
with a  point like object described by Kac-Moody spinor in order to take
into
accuunt the conformal degrees of freedom.  The cm coordinates of the
 boundary
components are kept fixed in the definition of the  quantum averaged
 spacetime and
these degrees of freedom must be quantized. The experience with QFT
 suggests
that one can second quantize boundary component Super Virasoro
representations
regarded as  quantum fields in
the interior of the convergence cube: second quantized  N-particle state
corresponds  to state of N topologically condensed small 3-surfaces with
boundary.\\
c) The action should be just the generalization of free KM action obtained
by minimal substitution to realize gauge invariances. Gauge field is defined
 by
N-S representation. Since  N-S representation contains bosonic KM Dirac
operator
and its conjugate one obtains polynomial action fourth order in operators
creating N-S states.  The difficult technical  problem is the construction
 of
vertices that is the action of N-S representation as a gauge potential on
 Ramond
representations and N-S representations.  \\
d) Classical background fields defined associated with effective spacetime
 surface
are also present and couple to the Kac Moody spinors. \\
 e)   The experience with the  p-adic counterpart of standard
model suggests  that in  boundary degrees of freedom single p-adic
convergence
cube  with size $L_p$ is   the natural  quantization volume. The propagators
 of
the   theory are  apart from additive constant equal to  Super
 Virasoro
generators $p^k\gamma_k-G_0+H^k\Gamma_k$ (fermions) and $p^2-L_0+h$ (bosons).

\vm

 One cannot exclude the  possibility is that the second quantization of
 Kac-Moody
spinors is directly related to the  p-adic nondeterminism for the dynamics
 of
boundary components. The world lines of boundary components are determined
by
second order differential equations, which allow integration constants
depending
on finite number of  positive pinary digits. For instance, p-adic geodesics
of
$M^4$ have $m^k$ and $dm^k/ds$ depending on finite number of positive pinary
digits.   This nondeterminism implies degeneracy of the  absolute minima of
effective
action in accordance with spin glass analogy.   A possible manner to treat
the situation is to use vacuum functional depending on world line. This
functional is analogous to the  action exponential for particle and should be
essentially equal to the action exponential associated with Kac-Moody spinor
 and
therefore generalized Dirac action.
 The
calculation of S-matrix elements necessitates averaging over the different
 vacuum
configurations and therefore one must perform summation over all possible
different world line configurations. Since world lines can split these
configurations  in fact are counterparts of Feynmann graphs. It would not
 be
surprising if this summation, which resembles functional integration would
not
give second quantized field theory for Kac Moody spinors.  It seems however
that this line of argument is not correct although the value of  the
effective
action for space time is not expected to depend much on the positions of
particle world lines.

\vm

Boundary degrees of freedom do not  give rise to graviton in the
scenario  explaining Higgs mechanism  and this suggests that
graviton  must correspond to interior degrees of freedom. The p-adic
 realization
of $Diff(M^4)$ invariance  indeed allows
massless states for the p-adic quantization volume and graviton and its
possible companions correspond to entire p-adic convergence cube or small
cube inside it (recall that localized p-adic plane waves are possible).
Graviton emission corresponds to the emission of small 3-surface via
topological
sum vertex and the exchange of graviton can occur between different p-adic
convergence cubes interpreted as particles such as hadron and leptons.
 The
general  rules  look  again quite obvious. Propagators are same as in
 previous
case. Vertices correspond to operators creating a state with the quantum
 numbers
of the emitted particle, say  graviton.   N-particle vertices are
essentially
time ordered n-point functions integrated over the interior of the p-adic
convergence cube and slashed between initial and final states.     The
vertices
describing the emission of several particles via topological sum vertex
 are
possible but probably the rates  processes involving several exchanges  are
extremely low since gravitational coupling characterizes the strength of
these
interactions.

\vm

The description of hadron reactions  necessarily
involves  topology changing processes for p-adic 3-surfaces and p-adic field
 theory
limit in cube $L_p$ is not all what is needed to describe these processes
 just as
perturbative QCD does not provide full understanding of hadronic reactions.
 An open
problem  is how to
construct the formalism describing  topological particle reactions in length
 scales
$L>L_p$.  Real physics probably  gives  a satisfactory    description for
 these
processes but in a more ambitious approach one should somehow glue  the
p-adic QFT:s
associated with neighbouring p-adic converge cubes together. One
possibility is to  construct p-adic manifolds from p-adic convergence cubes
 as
Feynmann diagram type objects. The lines  of these diagrams would correspond
to  the
sequences of convergence cubes glued together along their sides.  The sum
over the Feynmann diagrams would be reproducible from a second quantization
of some
field theory in the discrete  lattice formed by the  converge cubes the
vertices  of the theory being  defined in terms of n-point functions
associated with single convergence cube.
In  \cite{padTGD} the
concepts
of p-adic light cone and p-adic manifold (constructed by gluing together
 p-adic
converge cubes)  were introduced: in \cite{padTGD} it was suggested p-adic
 manifolds
might be the counterparts of Feynmann diagrams and a geometric  second
quantization
for  the p-adic conformal field theory was suggested.   In this formalism
single line of Feynmann diagram was supposed to be a part of so called
 p-adic light
cone. As already noticed  it seems that only the sequences of p-adic
convergence cubes
of size $L_p$ are needed and it is possible to perform second quantization
 in the
discrete lattice formed by these cubes.

\vm

p-Adic thermodynamics for boundary degrees of
freedom  seems to be necessary in order to get predictions
comparable  with experiments. p-Adic thermodynamics provides an
elegant  description of Higgs mechanism and consistency requires  the
thermal
averaging  of reaction rates, too.  It is not clear whether the p-adic
version
of the  standard model, which should provide the practical calculational
 tool, allows
to  use
 thermal masses as such. The technical problem is that p-adic square roots
 for
thermal $M^2$  appearing in Dirac equation need not be p-adically real
 number.
This problem is avoided if one uses Kac-Moody spinors truncated to spinors
 of
$H$ and introduces mass matrix $M^k\Gamma_k$ in $CP_2$ degrees of freedom
with
the property $M_kM^k=M^2$ with respect to Euclidian metric. This mass matrix
 is
just the counterpart of $CP_2$ contribution to the trace of the second
fundamental form.

\section{Symmetries of the theory}

In interior of $X^2$ the basic symmetries of the theory  broken
 include  $Diff(M^4)$, canonical transformations localizable
with respect to $M^4$ coordinates and containing local color transformations
as
a  subgroup.  These symmetries are broken by gravitational effects only.
 Point
particle limit for boundary components allows  besides conformal symmetries
and local canonical symmetries local
$so(4)\times u(1)$ broken down to electromagnetic gauge group and local
$so(3,1)$ broken down to local version of the isotropy group of 4-momentum
 of
world line.  All these symmetries posses super  extension: fermionic
 generators
are obtained by commuting the bosonic  generators with local $N=1$ super
symmetries and kappa symmetries defined by covariantly constant right handed
neutrino. Localization of symmetries is possible using arbitrary Fourier
expandable functions of $M^4$ coordinates and  $Diff(M^4)$ algebra.

\subsection{$N=1$ supersymmetry  and kappa symmetry }

In quantum TGD right handed neutrino generates $N=1$ supersymmetry.
  $N=1$
supersymmetry and  the related Kappa symmetry provide the means for extending
p-adic  conformal symmetry to super conformal symmetry: fermionic  Super
Virasoro generators are obtained  by  localizing  of $N=1$ super (Ramond
 case)
symmetry  or Kappa symmetry(N-S case).  The supersymmetrization of
local color and local canonical  transformations, Lorentz transformations
and
electroweak gauge transformations is achieved by taking anticommutators with
Super Virasoro algebra.  This is what is needed since in the papers
\cite{padmasses}
 devoted to p-adic Higgs mechanism p-adic super conformal
symmetry
together with super versions of color and other symmetries were  found to
 lead
to excellent predictions for particle mass spectrum.

\subsubsection{Definitions of $N=1$ super symmetry and kappa symmetry}

$N=1$ super symmetry generated by covariantly constant right handed neutrino
spinor has following form

 \begin{eqnarray}
\Psi &\rightarrow & \Psi+ \epsilon u_R\nonumber\\
 h^k&\rightarrow& h^k + \epsilon
\bar{u}_R \gamma^k\Psi+
 \bar{\epsilon} \bar{\Psi}\gamma^ku_R\nonumber\\
\
\end{eqnarray}

\noindent Here $\epsilon$ refers to the oscillator operator creating right
handed neutrino spinor with constant spatial dependence and $\bar{\epsilon}$
 is
the hermitian conjugate of this operator: the anticommutator of these
operators is proportional to unit operator.

\vm

The is  also a second supersymmetry  generated by right handed neutrino
analogous to the kappa symmetry of super string models \cite{Super}. This
symmetry is needed on boundary components only  to construct N-S type
representations and is defined as

\begin{eqnarray}
\Psi &\rightarrow & \Psi+ K u_R\nonumber\\
h^k&\rightarrow& h^k + \epsilon
\bar{u}_R \gamma^kK \Psi+
 \bar{\epsilon}
\bar{\Psi}K\gamma^{k}u_R\nonumber\\
K&=& n_k\gamma^k\nonumber\\
n^k &\equiv &\frac{D_{\alpha}\Gamma^{\alpha}}{ \sqrt{H^kH_k}}=
\frac{\gamma^kH_k}{\sqrt{H^kH_k}} \nonumber\\
\
\end{eqnarray}

\noindent    This supersymmetry doesn't
annihilate  on mass shell states as in
 string model.  The action of Dirac operator gives the trace of the second
fundamental form  and this implies analogy with kappa symmetry of string
model apart from the presence of normalization factor. Normalization factor
and therefore the direction of unit vector $n^k$  is
well defined unless  the boundary component in question  is minimal surface.
For boundary components idealized to world lines the value of
 $H^k\gamma_k$
is covariant constant defined by the nonvanishing vacuum weight of Kac Moody
spinor so that there are no problems.     Kappa  symmetry has the properties
needed  to  generate N-S type generators if one assumes that leptons move
in integer
modes and quarks move in half odd integer modes.

\subsubsection{Supersymmetrization of K\"ahler Dirac action}

$N=1$ supersymmetry suggests also a proper manner to generalize the K\"ahler
 Dirac action to p-adic context to obtain effective action.  Dirac action as
 such
transforms by total divergence in supersymmetry,  which does not act on
$H$-coordinates. One can however extend the super supersymmetry by replacing
$H$-coordinates with super coordinates:

\begin{eqnarray}
h^k&\rightarrow& h^k+ J^k\nonumber\\
J^k&=&  K\bar{\Psi}\Gamma^k\Psi\nonumber\\
K&=& K_0p^n,\ n>0
\end{eqnarray}

\noindent   in K\"ahler Dirac action.    This replacement breaks general
 coordinate invariance unless one fixes some physically preferred coordinates
for $H$.    In $M^4$ degrees
of
freedom the preferred coordinates are just the standard Minkowski
 coordinates.
 The simplest candidate  are complex coordinates with the property that the
 action
of $su(2)\times u(1)$ is linear. These coordinates are unique only apart
from color
rotation and internal consistency requires that physical states are color
singlets:
note that each p-adic regions is separately color singlets.
The nice feature of this choice is that color symmetry is exact and
therefore super
symmetrization works without any assumptions about the symmetries p-adic
 spacetime
surface.  That color confinement is related to super symmetry is in
 accordance with
the recent results  of  Seiberg and Witten \cite{Bigpaper}.

\vm

   The parameter   $K$ appearing in the defining formula for the super
coordinate  must be chosen to be proportional to some positive power of $p$
 in order to
obtain converging  perturbation theory.  The  convergence of the
perturbation theory is
extremely rapid for physically interesting values of $p$.  The deviation of
 the induced
metric  from flat metric must of order $K^2$.  For $K\propto p$ this means
that the
$CP_2$ part of induced metric is of order $O(p^2)$ as it should be.   Same
applies to the gradients of spinor fields.

\subsubsection{Why supersymmetry is so nice?}

Supersymmetrization is  physically well motivated  as the following
 considerations
show.\\
a) $Diff(M^4)$
symmetry implies that the oscillator operators associated with the
quantized fermion
fields create only electroweakly neutral massless states. Graviton is
contained in
the spectrum of the  massless states.   Each boundary component however
carries its own
quantized spinor fields and the total fermion number of 3-surface can at
most be  the
number of its boundary components. This means that elementary particle
 quantum numbers
 reside on boundary components  in topological condensate. This implies
 explanation of family replication phenomenon  in terms of boundary
topology excellent
predictions for lepton masses.\\ b) Supersymmetrization gives color
confinement as a
consistency condition: each p-adic region with fixed value of $p$ is color
 singlet. \\
c)
Elementary bosons other than graviton  correspond to boundary components.
 Boundary
component  spinor fields are indeed independent  degrees of freedom.
If $CP_2$
coordinates are supersymmetrized the  induced gauge potentials on  the
 boundaries  have
fermionic quantum parts and in  the lowest order gauge boson quanta can
be regarded as
fermion antifermion pairs:  this is just what physical consideration have
 forced to
conclude years ago.    The couplings of  the bosonic quanta are automatically
of correct form
and the interpretation of K\"ahler action as EYM action generalizes to the
  quantum
context.    It is easy to verify that the quantum contribution to the
induced
metric  is automatically of order $p$ so that gravitational interaction
strength  has correct order of magnitude automatically.  \\
  c) In TGD quark color triplets correspond to color partial waves,
which are expressions in $CP_2$ coordinates fixed uniquely by group
theoretical
considerations \cite{TGD}.  There  is however a  problem in flat $M^4$
 background
since $CP_2$ coordinates are constant. One possibility is to allow
 color partial
waves in $CP_2$ cm degrees of freedom of quark. Second possibility is
classical
description by fixing the classical color charges associated with the world
 line
of point like quark.  Supersymmetrization suggests a  third manner to
 overcome
the difficulty. The partial waves can  be constructed using purely quantal
$CP_2$ coordinates Ê$K\bar{\Psi}\Gamma^k\Psi$ and are clearly a purely
nonperturbative phenomenon!

\subsubsection{Can one  reproduce the Super Virasoro representations
associated
 with elementary particles?}

A basic test for the tentative  scenario is whether it can reproduce the
 general
 structure making possible the succesfull mass calculations performed in
 \cite{padmasses}.   The previous
 considerations
suggest that Kac Moody spinors correspond to representations of Super
Virasoro
transformations, which act on world line coordinate $t$ and on the complex
coordinate of
the boundary component so that  generators are simply sums of the
corresponding
generators and one can understand string mass formula.

\vm

 The rules for constructing Kac Moody spinors  are following.\\
a)  Single two-dimensional spinor corresponds to a p-adic counterpart of a
 unitary discrete series  $so(4)$ representation $(c,h)=(0,0)$
 representations
\cite{Goddard}.   Kac Moody counter part of $H$-spinor corresponds to 4-fold
tensor product of these representations, two representations in $M^4$
and two
representation in $CP_2$ degrees of freedom.  These representations give rise
to
Kac Moody version of Lorentz group $so(3,1)$ and $so(4)$.   To get Kac
 Moody
spinor fields one adds two additional tensor factors.  The group $U(1)$
associated with K\"ahler potential of $CP_2$  gives rise to additional
tensor
factor and color  $su(3)$ gives further tensor factor.   It turned
out that $su(3)$  is unique among Lie groups in that it  allows physically
acceptable
$(c,h)=(0,0)$ Super Virasoro representation.  \\ b) Gauge bosons correspond
 to N-S type
representations in each tensor factor.  Quarks correspond to Ramond
representations in each tensor factor. Leptons are exceptional in the sense
that
they correspond to N-S representations in $su(3)$ factor rather than being
Ramond type in every tensor factor. \\ c)
The super partners of various particles have quantum numbers of leptoquark
 in N-S
representations so that N-S type  super generators transform quarks into
leptons and
vice versa. In Ramond representations  super generators conserve fermion
number.

\vm

Is it possible to reproduce these structures in the proposed scnenario?  One
can
restrict the consideration to boundaries and  each boundary
 component  carries its own spinor fields satisfying its own Super Virasoro
conditions. \\ a) Right handed neutrino generates $N=1$ super symmetry and
respects fermion number conservation whereas kappa symmetry tranforms leptons
into quarks.    Localizing $N=1$ supersymmetry generators using half odd
integer or integer powers of $M$ one  obtains anticommuting Super Virasoro
generator in Ramond and N-S type representations respectively.
  Kappa  symmetry has the properties needed to
generate N-S type generators if one assumes that leptons move in integer
 modes
and quarks move in half odd integer modes obtained as square roots of
planewaves associated with leptons.    \\  b)
Although the action of Kappa symmetry  transforms quarks into right handed
neutrino and vice versa,  the states of N-S representations are eigenstates
of
baryon and lepton number: in particular,  bosons possess vanishing B and L.
The
point is that the state, which is  created by applying several operators
$F_{n+1/2}$ vanishes unless the oscillator operators associated with right
handed neutrino and its antiparticle anticommute pairwise to identity: this
in
turn implies that the quark number of the state vanishes.    G-parity rule
stating that only even or odd number of $F_{n+1/2}$:s appear in the operator
creating the state implies that all bosons have $B=L=0$. \\
c) One can construct local canonical transformations and as a special case
color transformations by multiplying the Hamiltonians with an analytic
function of $M^4$ coordinates. Super symmetrization of the algebra is
 achieved by
calculating the commutators with fermionic super Virasoro generators. \\
d) Local $so(4)\times u(1)$ is obtained  makes sense as broken symmetry,
when
boundary components are idealized with world lines and localization is
achieved by multiplying the generatores with some function basis
$f_n(t)=t^n$
world line coordinate. Again one must assume that leptons and quark type
 KM
spinors on world line  correspond to real p-adic plane waves and their
square
roots respectively.

\vm

One must keep in mind that the previous arguments say nothing about the
realization of super conformal or  $Diff(M^4)$  invariance in the  interior.
In
the interior  $Diff(M^4)$ seems to be a more appropriate symmetry and super
version of this symmetry for fermions probably involves only Ramond
representations
and  $N=1$ symmetry since the kappa symmetry is need not be well defined in
 interior
(the trace of the  second fundamental form is not covariantly constant and
 can
vanish).

\subsection{The action of super $Diff(M^4)$ (conformal transformations)}

$Diff(M^4)$ (conformal) transformation can be regarded as active
transformations in $M^4$ (boundary components).  One must replace the
ordinary coordinate
 $h^k$ with $h^k_S$ in the expression for the infinitesimal generators
 of the
$Diff(M^4)$ (conformal) transformation

\begin{eqnarray}
j^k(h)\frac{\partial }{\partial h^k } &\rightarrow &j^k(h_s)
\partial_k\nonumber\\ \partial_k&\equiv& \frac{\partial}{\partial h^k_S}
\end{eqnarray}

\noindent to guarantee that   $j^k(h_S)$ is invariant under super $N=1$
supersymmetry  and Kappa symmetry.
  The action of the  infinitesimal  $Diff(M^4)$ (conformal)  transformation
on $H$-spinors is  spin rotation

\begin{eqnarray}
\delta \Psi &=&   \epsilon\partial_k j_{l}\Sigma^{kl}\Psi \nonumber\\
\end{eqnarray}

\noindent   so   that fermions possess conformal spin $J=1/2$.

\vm

To derive the explicit form of Super $Diff(M^4)$ (Virasoro) generators one
 needs the
   explicit form of the
infinitesimal generators of  $Diff(M^4)$ (conformal) transformation,  super
symmetry and kappa symmetry as vector fields in the cartesian product
of $H$ and
spinor space

\begin{eqnarray}
J_C&=& j^k\partial_k + \delta \Psi \frac{\partial}{\partial \Psi}+
 \delta \bar{\Psi} \frac{\partial}{\partial \bar{\Psi}}\nonumber\\
J_F&=& j_F^k\frac{\partial}{\partial h^k}  +\epsilon O(F)
u_R\frac{\partial}{\partial \Psi}
 + \bar{\epsilon }\bar{u}_RO(F)\frac{\partial}{\partial \bar{\Psi}}
 \nonumber\\
j_F^k&=& \bar{\epsilon}\bar{u}_R\gamma^kO(F)\Psi +
\epsilon\bar{\Psi}O(F)\gamma^k u_R \nonumber\\ O(F=super)&=&1
\nonumber\\
O(F=kappa) &=& n^k\gamma_k \nonumber\\
n^k\gamma_k&\equiv& \frac{H^k\gamma_k}{\sqrt{H_kH^k}}\nonumber\\
Anti(\epsilon,\bar{\epsilon})&=&1
\end{eqnarray}

\noindent Here
 subscript $F=super/kappa$  refers to super symmetry or kappa symmetry
and $C$ to
$Diff(M^4)$ (conformal)  symmetry.

\vm

It is not at all clear whether kappa symmetry makes sense in interior
 whereas on the
boundaries kappa symmetry is necessary in order to reproduce the physical
 picture
suggested by the mass calculations.  The generator of  the might-be Kappa
 symmetry is
obtained by multiplying $u_R$ in the definition of super symmetry by
 $K= n_k\gamma^k$,
where $n_k$ is the unit vector field associated with the trace of the super
 symmetrized
second  fundamental form. For minimal surfaces    $n_k$ is ill defined.
 The
trace vanishes for the standard imbedding of $M^4$ and also  for the
 gravitationally
warped surfaces if the $CP_2$ projection of the surface is
geodesic circle.  On the boundaries situation is different.  When  boundary
 components
are idealized with world lines with the property that  $S^kS_k=M^2$
($S_k$ is the
bosonic part of the ordinary second fundamental form)   hold true, the unit
 vector
exists.  Note that   the existence of $1/\sqrt{H^kH_k}$ does not produce
problems if the
bosonic part of $H^k$  is nonvanishing. It will be found that  the value
of $H^kH_k$ on
boundaries is completely fixed.

\vm

 The simplest guess is that
super $Diff(M^4)$ (Virasoro) generators are obtained by suitably localizing
 $N=1$ super
symmetry and kappa invariance. The explicit form of the generators can be
deduced using following arguments. For simplicity the consideration is
 restricted to
conformal generators on boundary components.  The  generalization to
$Diff(M^4)$
generators acting on interior is obvious: the only difference is that
 both
quarks and leptons correspond to Ramond type representations in interior and
only
$N=1$ supersymmetry is needed.  \\ a)  The square of the  super generator
 is lightlike
translation $u_R\gamma^ku_R$ and therefore proportional to the  generator
$L_{-1}$ multiplied by suitable p-adic number in 4-dimensional extension.
Same
applies to kappa symmetry.  Hence the conformal weight of super and kappa
symmetry generators is $-1/2$. Super generator and kappa generator behave
essentially as $G_{-1/2}$ and it turns out that  the  anticommutators of the
local super/kappa symmetry generators
 generate the bosonic  parts of Virasoro generators.  Virasoro generators
however contain spin rotation term and therefore also local super generators
 must
contain analogous term, which must vanish for super generator and kappa
generator. A little consideration shows that  this term must act as vector
 field
in   $M^4$ degrees of freedom. \\
 b) The
simplest guess for  Ramond   generators  is following

\begin{eqnarray}
G_k&=& G^0_k +G_1^k\nonumber\\
G^0_k&=&  M_S^{k+1/2}J_{Super}  \nonumber\\
G^1_n&=& k_R(n) M_S^{1/2}(\bar{\epsilon}\bar{u_R}L_n \gamma^k\Psi +
 \epsilon \bar{\Psi} \gamma^k L_n u_R) \partial_k \nonumber\\
\
\end{eqnarray}

\noindent For N-S generators one has

\begin{eqnarray}
 G^0_{n-1/2}&=& M_S^nJ_{Kappa}  \nonumber\\
 G^1_{n-1/2} &=&  k_{N-S}(n)(\bar{\epsilon}\bar{u_R}KL_n
\gamma^k\Psi+
 \epsilon \bar{\Psi} \gamma^k L_nKu_R )\partial_k   \nonumber\\
K&=&\gamma^kn_k
\end{eqnarray}

\noindent where $M_S$ is complex $M^4$ super coordinate. The action of
$L_n$ on $u_R$ is spin rotation. The dependence of the parameter $k_R(n)$
($k_{N-S}(n)$)  on $n$ is fixed by the requierement that anticommutators of
the
 super
generators are correct.

\vm

To see that the ansatz works for a suitable choice of the parameters $k_n$
some
 calculation is needed. It is enough to study the
anticommutator of Ramond generators since  the calculation for N-S case is
essentially identical.\\
 a)  Super coordinate is annihilated by the
action of local  super symmetries $G^0_n$ so that only the derivative with
respect to $\Psi$ or $\bar{\Psi}$  appearing in the fermionic part
 appears in the
anticommutator of local supersymmetries $G^0_n$ and one has

\begin{eqnarray}
Anti(G^0_m,G^0_n)&=&
2M_S^{n+m}Anti(\bar{\epsilon},\epsilon) \bar{u_R}
\gamma^ku_R \partial_k = 2M_S^{n+m} \bar{u}_R\gamma^ku_R\partial_k
\propto L_{n+m}
\nonumber\\
\
\end{eqnarray}

\noindent The action of the light like translation on analytic functions is
equivalent with the action of $d/dM_S$ and one obtains apart from
multiplicative constant the bosonic part of the  Virasoro generator. Same
conclusion holds true in N-S case. \\
b) The action of the fermionic part of the  super symmetry $G^0_m$ on $G^1_n$
takes place via fermionic derivative and gives

\begin{eqnarray}
G^0_mG^1_n &+&(n\leftrightarrow m)\nonumber\\
&=&
k_R(n)M_S^{m+1/2}Anti(\bar{\epsilon}\epsilon)(
M_S^{1/2}\bar{u_R} \gamma^kL_n u_R \partial_k+(n\leftrightarrow m)
\nonumber\\
 &\propto& 2k_R m_S^{n+m} \bar{u}_R\gamma^ku_R\partial_k
\propto L_{n+m} \nonumber\\
\
\end{eqnarray}

\noindent If $k_R(n)$  is of form  $k_R(n)= k_R/(n+1)$,  this term is
also proportional to $L_{n+m}$ and   combines with the commutator of the
super
generators to give  $2L_{n+m}$ with  a proper choice of the constant $k_R$.
Same conclusion holds  true in  N-S case.  \\
 c)  Spin rotation
part of the Virasoro generator comes from the  action of the
  spin rotation term  $G^1_n$  on the fermionic part of local
 super-symmetry
$G^0_m$ (the action on bosonic part vanishes)  and  gives a term

\begin{eqnarray}
G^1_nG^0_m&+& (m\leftrightarrow n)\nonumber\\
&=&
k_R(n)M_S^{1/2}(\bar{\epsilon}\bar{u_R}L_n \gamma^k\Psi+
 \epsilon \bar{\Psi} \gamma^k L_n u_R)
\partial_kZ^{m+1/2}(\bar{\epsilon}\bar{u_R}\partial_{\bar{\Psi}}+\epsilon
u_R\partial_{\Psi}   )\nonumber\\
&\propto&  k_R(n)(m+1/2)
M_S^{m}(\bar{\epsilon}\bar{u_R}L_n \gamma^k\Psi +
 \epsilon \bar{\Psi} \gamma^k L_n u_R)(\bar{\epsilon}\bar{u_R}
\partial_{\bar{\Psi}}+\epsilon
u_R\partial_{\Psi}   ) + (n\leftrightarrow m)
\nonumber\\
\
\end{eqnarray}

\noindent For the  anticommutator to be proportional to  the spin rotation
 term
associated with $L_{n+m}$, it must be proportional  to $n+m+1$.
This is  the case if one has   (in accordance with the conclusion of  b)

\begin{eqnarray}
k_R(n)&=& \frac{k_R}{n+1}\ n\neq 0\nonumber\\
k_R(-1)&=&0
\end{eqnarray}

\noindent  For this choice  the anticommutator of  the oscillator operators
associated with covariantly constant right handed neutrino gives c-number
term and the action of derivative operator reduces to the action of the
light like
translation generator   and thefore the anticommutator is
proportional to the spin rotation term of $L_{m+n}$. With a proper choice of
$k_R$ and normalization of Super Virasoro generators it is possible to get
correct anticommutation relations. In N-S case one obtains similar result.
Note that  $n=-1$ does not produce trouble since spin rotation term vanishes
 in
this case and only the second  term in anticommutator contributes.

\vm

 The general expressions for $Diff(M^4)$  currents  can be
deduced from the general formulas   of  appendix.  The  contribution of
fermions to the conformal current    has besides the term
 proportional to energy momentum tensor also a spin term proportional to
$\bar{\Psi}\Gamma^{\alpha}\partial_lj_k\Sigma^{kl}\Psi$.        The  spin
rotation  term does not vanish.

\subsection{Local color symmetry and local canonical transformations
in interior}

Local canonical transformations are generated by Hamiltonians, which have
 arbitrary dependence
on $M^4$ coordinates

\begin{eqnarray}
H_{loc}&=& f(m^k) H
\end{eqnarray}

\noindent The change in the induced K\"ahler form is vanishing so that the
change in action comes from the change of induced metric and is very small.
Conformal canonical transformations are obtained by restricting the
functions $f$
to functions $M^n$.   The algebra contains local color transformations as a
subalgebra.

\vm

The action of the infinitesimal canonical transformation on the  induced
spinor
fields  has following form

 \begin{eqnarray}
J &=&  j^kA_k + \partial_k j_{l}\Sigma^{kl} +kH
\end{eqnarray}

\noindent In this expression  $h^k$ is replaced with $k^k_S$ everywhere.
 Here  the
covariant derivative contains also the $U(1)$ part necessitated by the
acceptable spinor structure of $H$  and $k$ is integer, whose value is
determined
by the coupling to K\"ahler potential and  proportional to fermion number,
which is one third for quarks.
 $H$ denotes the
Hamiltonian of the canonical transformation. The action of the   canonical
transformation on spinor connection implies additional terms in conserved
current,   which cancels the noncovariant piece proportional to $A_k$ in the
conserved current and only the spin rotation term and Hamiltonian term are
left.
 Hamiltonian term. The variation of $\Gamma_{\alpha}$ gives also a
contribution
to  the  current.

\vm

Canonical transformations generalize to supercanonical transformations just
as
conformal transformations do  and one obtains Ramond and N-S type
representations of super canonical symmetries. The currents associated with
 super canonical transformations are obtained by
applying the general receipes of Appendix.

\subsection{Realization of $Diff(M^4)$ invariance allows graviton.}

$Diff(M^4)$  tranformations   and their  super counterparts  act as
approximate symmetries of the theory in the interior of the spacetime
surface. This
means that the approximate symmetry group is even larger than the group of
p-adic conformal symmetries and that the theory is analogous to topological
field theory.  This is rather satisfactory feature since it gives final
justification for the hypothesis that partons correspond to boundary
components.
The result however raises the question whether the spectrum in the interior
 contains
any $Diff(M^4)$ invariant states besides  vacuum state.  Physical intuition
suggests
that graviton and possibly some massless partners of the  graviton should
 be
contained  in the spectrum (after all, the starting point of TGD was to
 construct
Poincare invariant theory of gravitation!)   The problem is to understand
how
graviton can circumvent the extremely tight constraint of $Diff(M^4)$
invariance.
 Standard anticommutation relations for fermions seem to imply
essentially
topological field theory.  It turns out that  p-adic massless of fermions
allow
to circumvent the difficulty.

\vm

The point is that p-adic planewaves with positive  p-adic momentum $k$ are
 only
needed and $k\rightarrow -k$ defines complex conjugate and functions
expressible as Fourier series of negative momenta belong to the dual of
the function   space. This implies that fermionic creation operators
correspond
to momenta $k$ with each $k_i>0$ and possessing finite number of positive
 pinary
digts.   The following modified definition of fermion vacuum so that time
 and
spatial directions are treated in same manner

\begin{eqnarray}
a_n\vert vac\rangle &=&0,  n_i<0 \ for \  some \ i
\end{eqnarray}

\noindent The formula states
that oscillator operator acts as creation operator only provided all
 components
of the four momentum are positive.  The
anticommutators of the  fermionic oscillator operators

\begin{eqnarray}
Anti(A_m,A_{-n}) &=&\delta (m-n)
\end{eqnarray}

\noindent  are of the standard form.

\vm

The commutation relations of $Diff(M^4)$ generators in p-adic planewave
 basis
 $J(\epsilon, k)$, where $\epsilon$  is polarization vector characterizing
the
direction of flow induced by generator and $k$ with $k_i>0$ for each $i$,
is
momentum of the p-adic  planewave read as

\begin{eqnarray}
Comm(J(\epsilon_p,p_1),J(\epsilon_2,p_2))&=& \epsilon_1\cdot p_2
J(\epsilon_2,p_1+p_2)- \epsilon_2\cdot p_1 J(\epsilon_1,p_1+p_2)\nonumber\\
\
\end{eqnarray}

\noindent On any light like geodesic $\epsilon\cdot \epsilon=0$,  with $p$
and
 $\epsilon$ parallel,   these commutation relations   become
trivial. The central extension  parameter $c$ for $Diff(M^4)$ are expected
 to be
vanishing since the only source of c-number terms is normal ordering of the
 fermion
generators in the expressions for translation generators $J(\epsilon,0)$.
 In
the lowest order the only possible source of normal ordering terms is spin
rotation
term, which however vanishes identically for translations.    One can again
define $Diff(M^4)$ generators associated with negative p-adic momenta as
 Hermitian
conjugates of the positive momentum generators and they act in the dual
of the function
space.

\vm

  $Diff(M^4)$
 generators $J(\epsilon, -n)$ and other gauge group generators  must
annihilate
physical states if $n$ is nonzero and $n_i$ are nonnegative:

\begin{eqnarray}
J(\epsilon,-n)\vert phys\rangle&=&0 \ n_i\ge 0, \ n\neq 0
\end{eqnarray}

\noindent and it turns out that these conditions allow massless states.
Translations  correspond to generators  $J(\epsilon, 0)$ and  these
 generators
cannot annihilate nonvacuum physical states. That complete $Diff(M^4)$
invariance
is not required is in accordance with the fact that  $Diff(M^4)$
invariance broken by gravitation.

\vm

Consider now the action of $Diff(M^4)$ generators  $J_{\epsilon,-n}$
 $n_i\ge 0$
 on a  state created by a  fermionic creation  operator with
light like   momentum $p$ with positive components.  The action of
$Diff(M^4)$
generator  on  the oscillator operator is induced from its action on
 planewave with
momentum $p$ and the action must be  of the  form

\begin{eqnarray}
Comm(J(\epsilon,-n),a_p)&\propto& \epsilon\cdot p  a_{p-n}
\end{eqnarray}

\noindent The action  vanishes if $ \epsilon$ and $p$ are parallel or
 $\epsilon$ lies in transversal polarization space to $p$.   The action
gives annihilation type operator,  when $n$ has momentum component
$n_{perp}$ in
transversal polarization space of $p$ since $n_{perp}$ has positive
 components
so that states created using only oscillator operators with momenta in
direction of $p$ are automatically annihilated by these $Diff(M^4)$
generators.
When $\epsilon$ is in light like direction conjugate to $p$ and $n$ is in
two-dimensional Minkowskian place spanned by $p$ and its light like
 conjugate
the action is nonvanishing. The resulting oscillator operator is creation
type
operator  for $p_3-n_3\ge 0$  so that one obtains finite number of
 conditions.
Actually the conditions are equivalent with the ordinary Virasoro
conditions.

\vm

The  representation in question is reducible
since there there are several types of fermionic oscillator operators
labeled by their electroweak quantum numbers.  This means that
nontrivial solutions of the  Virasoro conditions are obtained and the
 number
of the solutions is actually infinite.  Therefore there are good hopes of
obtaining graviton and its companions as $Diff(M^4)$ invariant states.
 Additional
constraints come from Super Diff generators.

\vm

There are also additional conditions coming from the  canonical invariance
 of the
states. Canonical transformations induce $CP_2$  spin rotation and $U(1)$
tranformation on the states and this means that canonical invariants must be
constructed as Fourier components with momentum $p$ of spin vector currents
  $\bar{\Psi}\gamma^k\Psi$ invariant under $CP_2$ spin rotations and
$U(1)$ rotations.$\Psi$ can have either  lepton or quark type chirality.

\begin{eqnarray}
B(L,p,\epsilon)&=&  \bar{L} EL (p)\nonumber\\
B(q,p,\epsilon) & =& \bar{q} Eq (p)\nonumber\\
E&=& \epsilon_k\gamma^k\nonumber\\
E\cdot p&=&0
\end{eqnarray}

\noindent  Here
$\gamma^k$ is $M^4$ gamma matrix  and creation operators corresponds to
either
quark or lepton chirality: spin  indices are written explicitely.  Canonical
invariance  drops from consideration massless one-fermion states and only
 integer
spin states with vanishing B and L remain under consideration. The
quantities
$B(p,\epsilon)$ can consists of quark or lepton type creation operators.

\vm

The constraints coming from Super symmetry are satisfied if  one drops
from $\bar{\Psi}\gamma^k\Psi$ away the part quadratic in oscillator operators
 associated with
right handed neutrino and replaces $M^4$ coordinate with its super
counterpart
in p-adic planewes so that p-adic planewaves become supersymmetric. In a
 concise
form one can express the invariants $B$ as

\begin{eqnarray}
B(L,p,\epsilon)&=&  \bar{L}EL(p)-\bar{\nu}_RE \nu_R \nonumber\\
B(q,p,\epsilon) &=& \bar{q}Eq(p) \nonumber\\
p\cdot \epsilon&=&0
\end{eqnarray}

\noindent Only two transversal polarization vectors are needed. The
 operators
$B(F,p,\epsilon)$ commute with each other and the  commutators of
 $Diff(M^4)$
generators with operators $B$ are proportional to operators $B$ themselves
 and
 Diff invariance conditions for state with momentum $p=n$
reduce to Virasoro conditions for $L_k$,$ k=1,...n$.
This implies that
 $Diff(M^4)$ invariant states form algebra and the problem is to identify the
generators of this algebra as
monomials of operators $B$. The generators are labeled by the numbers
 $N(L)$
and $N(B)$  of $B(L,..)$ and $B(q,..)$ in the monomial,  by polarization
 tensor
 $\epsilon_{kl...n}$  in transversal degrees of freedom (spin) and by the
total
momentum $p$. A possible  interpretation of the generators is as elementary
bosons and the products of generators as many boson states with collinear
momenta. If this interpretation is adopted one obtains dilaton,
antisymmetric tensor ($J=0$) and graviton as elementary bosons. Higher
spin states  can perhaps  be regarded as many boson states.

\vm

\noindent a) $N=1$ states are just zero momentum spin one states created by
$B(F,\epsilon,0)$ , $F=L,q$. \\
 b)  $N=2$  states  are of type $B(L)B(L)$, $B(q)B(q)$ and
$B(L)B(q)$.
For $p>2$ states Virasoro conditions allow no solutions and only possible
states have  $p=0$ or  $p=1$.    $N=2$ states with zero momentum are obtained
as products of spin 1 generators: also spin two is possible.  $p=1$ states
  are
possible, when zero modes are allowed. The states are  of  the form

\begin{eqnarray}
\sum_{\epsilon_i} c_{\epsilon_1,\epsilon_2}
(B(F_1,\epsilon_1,p=1)B(F_2,\epsilon_2,0)-B(F_2,\epsilon_1,1)
B(F_1,\epsilon_2,0))
 \end{eqnarray}

\noindent   $L_1$ Virasoro condition is satisfied by antisymmetry under
exchange $F_1\leftrightarrow F_2$. $B(L)B(q)$ states  allow
dilaton,antisymmetric tensor ($J=0$) and graviton.  For $B(F)B(F)$  only
antisymmetric tensor state  is possible. \\
c) For $N=3$  one obtains  $p=0$ and $p=1$  states constructed as products
 of
$N=1$ and $N=2$ states: the largest spin of this kind of state is $J=3$. It
seems that no
 $N=3, p=1$ generating states  exist. Spin one state with nonvanishing
momentum
is possible and this is due to the fact that only two polarization vectors
are
at use.\\ d) For $N=4$ one obtains products of $p=0,1$ generating states.
 Spin
$J=4$ is possible for two graviton state. Also $p=2$ graviton and scalar
 state
containing $B(L,\epsilon, p=2)$ is possible  in leptonic sector and is
constructed using the operator
$B(L,\epsilon,p=2)B^2(L,\epsilon_+,0)B(L,\epsilon_-,0)$. This state is
obtained as a special case of a more general construction procedure.

\vm

The states constructed do not include any higher spin exotic states.
 States  with arbitrarily large  values  of $N$ and $p$  are possible and
state space has natural shell structure
as the following arguments show.\\
i) To construct a state containing  $B(L,\epsilon, p=2)$ operators construct
first a zero momentum vacuum using the operator
$B(L,\epsilon_+,0)^2BB(L,\epsilon_-,0)$. This state has spin $J=1$ since the
lacking right handed neutrino pair acts as a spin 1 hole. This state is
annihilated by leptonic creation operators with $p=0$. For quarks
corresponding
state is created by  $B^2(q,\epsilon_+,0)B^2(q,\epsilon_-,0)$ and
has vanishing
spin. This state satisfies gauge conditions and is
 analogous to full electron shell in atomic physics. Apply to this state
operators $B(F,\epsilon,2)$. The resulting states satisfy gauge conditions
automatically by fermion statistics.\\
 ii)    The procedure can be continued. Fill first $p=1$ shell
using using $3$ ($4$) leptonic (quark like) operators $B(F,\epsilon,p=2=n)$.
By
commutativity  of $B$ operators only the $p=1$ fermionic creation operators
are
effective in this state.  The new vacuum has spin $J=2$($J=0$) for $F=L$
$(F=q)$
and applying $B(F,\epsilon,p=4=2n)$ operators one obtains a new shell
containing
$p\leq 2$ fermionic creation operators.\\
iii)  It it is obvious that the procedure can be
continued indefinitely and one obtains infinite number of states. The
resulting
states can be regarded as many particle states if full shell is regarded
 as a
new vacuum state. The vacuum states in turn  can be regarded as many
boson states constructed from previous vacuum state and states are just
products of  leptonic and and quark type states unlike the states obtained
directly from the vacuum.  This suggests that only the
  $N=2,p=1$ generators
should be regarded as operators creating single  particle states.  The
peculiar
consequence of the supersymmetry is the possibility of states possessing
arbitrarily large spin consisting of $J=1$ holes made from pairs of right
handed neutrinos.

\subsection{Local $so(4)\times u(1)$ is broken in standard manner}

Kac-Moodyt spinors used in  the mass calculations are representations of the
local
 $so(4)$
and $u(1)$ Kac Moody algebras and Quantum TGD as well as the success of the
  mass
calculations suggests  that   local $so(4)\times u(1)$ do not act as gauge
symmetry of the  K\"ahler Dirac action.

\vm

 The
natural identification of $so(4)\times u(1)$ is as the gauge group of $CP_2$
spinor connection.  Global $U(1)$ is exact symmetry.
$so(4)$ rotations are exact symmetries only  for the surfaces
possessing one-dimensional $CP_2$  projection implying vanishing of the
induced
gauge fields.  This condition is  in general not true.   Condition however
holds true, when the orbits of the  boundary components are idealized with
world
lines so that global $so(4)$ is good approximate symmetry at parton level,
whereas it is broken in the interior of the  spacetime surface.

\vm

The localized $so(4)\times u(1)$ transformations are not exact symmetries in
the point particle limit.  The change of Dirac action  density in a
local gauge  transformation reduces to  a total divergence by the
 conservation of
the charges associated with the global transformations if the the trace of
the
second fundamental form vanishes: this is not however the case.

\begin{eqnarray}
\delta L_D &=& D_t(\bar{\Psi}Anti(\Gamma^{t},Q)\Psi f_n(t))-X\nonumber\\
X&=&  D_t(\bar{\Psi}Comm(\Gamma^t, Q)D_t\Psi)f_n(t)  \nonumber\\
 f_n(t)&=&t^n
\end{eqnarray}

\noindent  Here $Q$ is the generator of the local gauge transformation.
 Nonconservation  term in general doesn't vanish: it differs from
conserved charge in that anticommutator of $A$ and $\Gamma^t$ is replaced
 with
commutator and the commutator does not in general vanish.

\vm

 The term $X$ characterizing local  charge nonconservation leads
to  the  standard form of electroweak symmetry breaking.   For left handed
charges $Q$  ($I^1_L$ and $I^2_L$ associated $W^{\pm}$ bosons)  it is
impossible
to find $\Gamma_t$ commuting with $Q$. For either purely vectorial or axial
generator  it is possible to satisfy the commutatitivity condition.
Electromagnetic charge is the only purely vectorial electroweak  charge
and the
vectorial  generator $\Sigma_{12}$ appearing in electromagnetic charge
commutes
with $\Gamma^t$ when the condition

\begin{eqnarray}
\Gamma^t&=& a(t)\gamma^0 +b(t)\gamma^3
\end{eqnarray}

\noindent  is satisfied: here the  standard definition of $CP_2$ vielbein
is
used \cite{TGD}. The commutativity of $\Gamma^t$ with $Q_{em}$ can be taken
 as
additional condition on the time development for the world line of the
 boundary
component in $CP_2$ degrees of freedom: this condition is equivalent with
 the
conditions stating that the projection of the  worldline is orthogonal to
 $e^A$,
$A=1,2$:

\begin{eqnarray}
e^{A}_k \frac{ds^k}{dt}&=&0, \  A= 1,2
\end{eqnarray}

\noindent Here $e^{A}_k$ are the components of the vielbein vectors  of
 $CP_2$ and
implies that induced $W^{\pm}$ fields vanish on the world line. This
 guarantees
that different charged states for spinor field are not mixed and
electromagnetic charge conservation results.
 The conditions imply that in standard coordinates
$(r,\Theta,\Phi,\Psi)$ (\cite{TGD}) the coordinates $\Theta$ and $\Phi$ are
constant on the world line of the particle.  One must however notice that
p-adic differentiability allows $\Theta$ and $\Phi$ to depend on finite
number
 of positive pinary digits of time variable so that these variables are
constant
only below some time  scale.

\vm

It is interesting to examine this condition, when $M^4$ projection of
world line
is geodesic line.  Combining the  condition
with the requirement that $S^k$ is covariantly constant along world line
(this
guarantees   the condition  $S_kS_k=M^2$)   one obtains
the conditions

\begin{eqnarray}
e^0_rS^r&=& A \nonumber\\
e^3_{\Psi}S^{\Psi}&=&B\nonumber\\
A^2+B^2&=&M^2
\end{eqnarray}

\noindent where $A$ and $B$ are p-adic constants depending on finite number
 of
pinary digits of the time coordinate. The resulting conditions give two
second
order differential equations with 4 p-adic integration constants
corresponding
to fixing arbitrarily finite number of pinary digits for  $(r,\Psi) $ and
$(dr/dt,d\Psi/dt)$ as a function of $t$.  One must notice that
conservation laws give constraints, which reduce p-adic nondeterminism. For
instance, p-adic geodesic lines of $M^4$ should look like ordinary geodesic
lines. Furthermore, if p-adic derivative is algebraic operation as it is
for p-adic
planewaves p-adic nondeterminism is lost.

 \subsection{Lorentz invariance and local Lorentz transformations}

  The first problem related to the realization of p-adic Lorentz invariance
is  whether the p-adic convergence cube for square root function is
invariant under Lorentz transformations or its subgroup.   Real word
 geometric intuition suggests
that this is   cannot be the case.  p-Adic Lorentz group
is however  quite different from its real counterpart. The existence of
the  exponent for
Lie-algebra element requires that the element is proportional to $\sqrt{p}$
at
least.  This means that the real conterparts of Lorentz transformations
 related continuously to identify are
small and these transformations form a group.   If the
 convergence cube has size  $L_p$ then Lorentz
transformation with  generators are proportional to $ \sqrt{p}$ at least
 leave the convergence cube invariant.

\vm

Momentum spectrum is not invariant under  small Lorentz transformations.
 p-Adic convergence cube implies infrared cutoff in momentum: choosing scale
suitably  suitable  the momentum components have  the expansion

\begin{eqnarray}
P&=&  \sum_{k>1}P_kp^{-k}
\end{eqnarray}

\noindent  The cutoff is obviously particle in the box effect.
  Small Lorentz transformations do not leave this expansion invariant since
they
induce non allowed p-adic digits to momentum components: the  contributions
 of
these digits to the real counterparts of momenta are  extremely small.
 Perhaps
one must just  accept the noninvariance of the momentum spectrum under
Lorentz
transformations as  a physical fact.  The lowest order
contribution to the energy is given by rest mass and is of order
 $O(p^{1/2})$
and Lorentz invariant  since
small Lorentz transformations induce only  $O(p^{3/2})$ correction to
 energy.

\vm

One must emphasize that there are also 'large'  Lorentz transformations
for
 which nondiagonal  matrix elements are of order $O(p^0)$.   The
orthogonoality
conditions for the matrix elements of these tranformations give strong number
theoretic constraints on the matrix elements of these tranformations. An
especially interesting subgroup  of Lorentz group is rational
Lorentz-transformations.   It is impossible to generate them using
exponential map since exponent function does not converge for Lie-algebra
element having p-adic norm  one. It would be interesting to understand the
detailed
 physical consequences deriving
from  the factorization of the  p-adic  Lorentz transformations  into a
 product of
large and small transformations. Perhaps the so called  nonrelativistic
regime
 corresponds to small Lorentz transformations leaving p-adic convergence
 cube
invariant.

\vm

Local  $so(3,1)$ appears also in the context of Kac-Moody spinors and one
 can
wonder whether these tranformations are local symmetries for
boundary-components in the  point like limit. Exactly the same arguments as
 before
can be carried out as for local $so(4)$ to find that those local Lorentz
transformations for which the charge generator $Q= q_{kl}\Sigma^{kl}$
commutes with $\Gamma^t$ define  local worldline symmetries broken by
 gravitation
only. Not suprisingly, this group is  the isotropy group of four-momentum
and for
p-adic
geodesic lines the group in question is independent of $t$ in sufficiently
short time scales.

 \section{  K\"ahler Dirac  action and general field equations }

In the sequel K\"ahler action and field equations will be discussed in a
 more
detailed form.

\subsection{K\"ahler Dirac action}

K\"ahler action density in the interior  is given by

\begin{eqnarray}
L_K &=& \frac{1}{4g_k^2} J^{\mu\nu}J_{\mu\nu}\sqrt{g}
\end{eqnarray}

\noindent $g_k^2$  is p-adic version of K\"ahler coupling strength. The
 experience with field theory limit of TGD suggests that p-adic K\"ahler
coupling strength is related to the real coupling strength via the
relatiohinship

\begin{eqnarray}
g_k^2&=&g_K^2p
\end{eqnarray}

\noindent In this manner the real counterpart of  the p-adic  K\"ahler
 coupling
strength does not deviate too much from its real value.
K\"ahler action can contain also a boundary term, which is Chern Simons
type
 term, which we do not bother to write explicitely here.

\vm

 Dirac action density  is given by

\begin{eqnarray}
L_D &=& (\bar{\Psi} \Gamma^{\alpha}D^{\rightarrow}_{\alpha}\Psi - \bar{\Psi}
D^{\leftarrow}_{\alpha}\Gamma^{\alpha}\Psi )\sqrt{g}\nonumber\\
 \Gamma_{\alpha}&=& \Gamma_k
\partial_{\alpha}h^k \end{eqnarray}

\noindent and differs from the ordinary Dirac action in that gamma matrices
are  dynamical variables  obtained by inducing the gamma matrices of $H$ to
4-surface. In \cite{TGD} it was suggested that induced gamma matrices
describe
the presence of fermions but it seems that the proper interpretation is as
a new
bosonic field.    Dirac action contains also parts associated with the
 boundaries of $X^4$ and each boundary component carries its own spinor
field
$\Xi$ so that boundaries become carriers of fermion numbers: this is the
 only
manner to achieve conservation of fermion  number currents in Quantum TGD.
 The
Dirac action for boundary component has exactly the same form as the
 interior
action and we do not bother to write it explicitely.

\vm

Super symmetrization of K\"ahler Dirac action is achieved by adding Super
 symmetric part to the bosonic coordinates.   This is in accordance with
 general
coordinate invariance if physically preferred coordinates are chosen for $H$.
The coordinates in question are standard Minkowski coordinates for $M^4$
and
complex  coordinates of $CP_2$ transforming linearly under
$su(2)\times u(1)$:
as a by product color singletness of physical states emerges as consistency
requirement.

\subsection{Field equations in the interior}

 Field equations are interpreted
as field equations for the effective action defining fermionic field theory
in
average space time associated with a given quantum state. This implies the
needed back-reaction from matter to geometry. In particular, for diagonal
 matrix
elements of  bosonic field equations one obtains different fermionic source
term
for each state and background spacetime depends on the quantum numbers of
 the
state.  It is convenient to allow bosonic solution to depend on maximal
number
of commuting operators defining the quantum numbers of the state. In
particular,
background spacetime depends on various YM charges and on the four-momentum
of
the state.  If boundary components are treated as point like objects,  the
square for the trace of  the  $CP_2$ part  the second fundamental form gives
modular contribution to the  mass squared of fermion.

\vm

It seems natural to solve the field equations by starting from the  standard
imbedding of $M^4$, to solve fermionic Dirac equation and then calculate the
change of the  space time metric (or deviation of   $CP_2$ coordinates from
constancy) by requiring that bosonic field equation remains true in given
 state
with momentum $P^k$.  Gravitational warping is in some sense the lowest order
backreaction effect and depends on the four-momentum of the state, only.
 This
effect is necessary in order to guarantee well definedness of perturbation
series.

\vm

The general field equations of
 the
theory are obtained by varying the  action with respect to $h^k$ and  $\Psi$.
 It is convenient to use $h^k_S,h^k
$  and $\Psi$ as basic variables and
introduces the definition of super coordinate as a constraint to the action
density:

\begin{eqnarray}
L =L_K+L_D+L_{const}&= & L+ \Lambda_k(h^k_S-h^k- K \bar{\Psi}
\Gamma^k (h^m_S)\Psi)
\end{eqnarray}

\noindent  The field equations are obtained by varying  with respect to
 $\Lambda, h^k_S,h^k$ and $\Psi$.

\vm

Consider first the  variation with respect to $h^k_S$ in the interior: the
 field equations are obtained
 from the field equations of nonsupersymmetrized action by replacing $h^k$
 with
$h^k_S$ and by adding constraing term $\Lambda_k$ to the right hand side.

\begin{eqnarray}
\frac{\delta S}{\delta h^k (x)}&=& \partial_{\alpha}(\frac{\partial
(L_K+L_D)}{\partial ( \partial_{\alpha}h^k_S}))-
  \frac{\partial (L_K+L_D)}{\partial h^k_S} -F_k\nonumber\\
F_k&=& \Lambda_k- K\Lambda_l\bar{\Psi} Comm(V_k,\Gamma^l)\Psi\nonumber\\
Comm(A,B)&\equiv&AB-BA
\end{eqnarray}

\noindent   The functional derivative of K\"ahler action with respect
 to $h^k(x)$ is

\begin{eqnarray}
\frac{\delta S_K}{\delta h^k (x)}&=&  D_{\alpha}(T^{\alpha\beta}
 \partial_{\beta}h^lh_{kl}) - j^{\alpha} J_{k l } \partial _{\alpha}h^l
\nonumber\\ T^{\alpha\beta}&=& \frac{1}{4g_K^2}(J^{\nu\alpha}J_{\nu}^{ \
\beta}-
g^{\alpha \beta} J^{\mu \nu} J_{\mu \nu}/4)\sqrt{g}\nonumber\\
j^{\alpha}&\equiv& D_{\beta}J^{\alpha\beta}
\end{eqnarray}

\noindent  The covariant derivatives appering in the various expressions
contain both the ordinary Riemannian connection acting on $X^4$ indices and
the
Riemannian connection of $H$ acting on $H$ indices.

\vm

 The functional derivatives of Dirac action with respect to H-coordinates
are

\begin{eqnarray}
\frac{\delta S_D}{\delta h^k (x)}&=&
D_{\alpha}(T_D^{\alpha\beta }\partial_{\beta}h^k +
 \bar{\Psi}\Gamma_kD^{\alpha}\Psi )
+  \bar{\Psi} \Gamma^{\alpha} F^k_l \partial_{\alpha} h^l\Psi\nonumber\\
T_D^{\alpha\beta}&=& \bar{\Psi}( \Gamma^{\alpha} D^{\beta}+
 \Gamma^{\beta}D^{\alpha})\Psi \sqrt{g}\nonumber\\
\
\end{eqnarray}

\noindent    Divergence of the energy momentum tensor comes from the
 variation
 of metric and term proportional to $\Gamma^k$ comes from the variation of
induced gamma matrix.  The term proportional to spinor curvature $F_{kl}$
 comes
from the variation of  spinor connection and divergence of energy momentum
current comes from the variation of metric.

 \vm

 The resulting interior field
equations have the character of hydrodynamical conservation equations.
 Only
four of them are independent by general coordinate invariance and for
 surfaces
reprsentable as maps $M^4\rightarrow CP_2$ the equations can be replaced
with
equations stating conservation of four-momentum density obtained by
 contracting
the canonical energy momentum tensor $T^{\alpha\beta}$ of K\"ahler
 Dirac action
with the gradients $\partial_{\beta}m^k$.   Probably the situation is same
 now
since $\Lambda^k$ vanishes for $M^4$ indices.

\vm

The  variation with respect to
$h^k$ affects only the constraint term and gives

\begin{eqnarray}
\Lambda_k&=&0
\end{eqnarray}

\noindent so that the constraint force $\Lambda^k$ vanishes in both
 $M^4$ and $CP_2$ degrees of freedom and field equations associated with $H$
coordinates are just standard equations with $h^k$ replaced with $h^k_S$.

\vm

The  variation with respect to $\Psi$ gives the Dirac equation for the
 induced
spinors in interior with $h^k\rightarrow h^k_S$  replacement  and additional
 source term
coming from the variation of the constraint equation

\begin{eqnarray}
(\Gamma^{\alpha}D_{\alpha} +\frac{1}{2}\Gamma_k H^k)\Psi&=&
K\Lambda_k \Gamma^k \Psi=0
\end{eqnarray}

\noindent  Here $H^k$ is the super symmetric generalization of the second
 fundamental form

\begin{eqnarray}
H^k_S&=& g^{\alpha\beta}D_{\beta} \partial_{\beta}h^k_S
\end{eqnarray}

\noindent  The presence of second fundamental form implies mixing of $M^4$
 chiralities.  The
constraint force term vanishes  by previous results so that field equations
 are
of same form as in standard theory.

\vm

 The vanishing of $\Lambda^1_k$  implies   complete decoupling of K\"ahler
 and Dirac actions as also happens in Quantum TGD.  One just takes the
 absolute
minimum of ordinary p-adic K\"ahler action and replaces the $H$ coordinates
 with
their supersymmetrized versions with $\Psi$  solved iteratively from Dirac
equation.

\vm

The equations for H-coordinates must be posed   as conditions on
physical states analogous  to  and probably equivalent with Virasoro
conditions.
  This means that Super Virasoro condition could well be an approximation to
the actual field equations true only in the limit that $CP_2$ metric
gives no
contributions to the induced metric.  Same conclusion  applies to canonical
invariance and its super counterpart.

\subsection{Field equations on boundaries}

The field equations on boundaries  can be derived straightforward manner.
 \\
a) Dirac equation on boundary reads as

\begin{eqnarray}
(\Gamma^{\alpha}D_{\alpha} +\frac{1}{2}\Gamma_k H^k)\xi&=& \Gamma^n\Psi
\end{eqnarray}

\noindent Interior spinors act as a  source term for boundary spinors. \\
b)  Field equations for $h^k(x)$ state that the functional derivatives of
 total action with respect to $h^k(x)$ vanish on boundaries. The functional
derivative of total  K\"ahler action  including Chern Simons term on
boundary
is given by

\begin{eqnarray}
\frac{\partial S_K}{\delta h^k (x)}&=& T^{n\beta} \partial_{\beta}h^k  -
(J^{n\beta}-J_*^{n\beta})J^k_{ \ l }
\partial_{\beta}h^l
\end{eqnarray}

\noindent  $J_*$ denotes the dual of induced $CP_2$ K\"ahler form.  It is
 assumed that the ratio of the coefficients
$k_1$ and $k_2$  multiplying Maxwell action and Chern Simons term
respectively in the definition of K\"ahler function equals to
$k_1/k_2=1$  (for various definitions see \cite{TGD}).  \\
c) The functional derivative of the  Dirac action with respect to $h^k(x)$ on
boundary is given by

\begin{eqnarray}
\frac{\partial S_D}{\delta h^k (x)}&=&
D_{\alpha}(T_{DB}^{\alpha\beta }\partial_{\beta}h^k) +
\bar{\xi}\Gamma_kD^{\alpha}\xi ) \nonumber\\
 &+& \bar{\xi} \Gamma^{\alpha}F^k_l \partial_{\alpha} h^l\xi
+T_D^{n\beta}\partial_{\beta}h^k \nonumber\\
 T_{DB}^{\alpha\beta}&=& \bar{\xi}( \Gamma^{\alpha} D^{\beta}+
\Gamma^{\beta}D^{\alpha})\xi \sqrt{g_3}  \nonumber\\ \
\end{eqnarray}

\noindent The index  $B$ refers to boundary in the above expressions.
$F_{kl}$ denotes the components of spinor curvature.

\vm

These equations have counterparts for Kac Moody spinors and should take
care of
conservation laws besides giving the KM Dirac equation.

\section{General solution of field equations}

\subsection{Perturbative solution}

  The general solution of the field equations  is obtained by expanding
  $\Psi$ in powers of $K\propto p$

\begin{eqnarray}
\Psi&=&\sum_n K^n\Psi_n
\end{eqnarray}

\noindent and solve $\Psi_n$  iteratively power by power.  Since
the  super part of $H$ coordinate is proportional to
 $K\propto p$,
one indeed obtains rapidly converging power series in $K$.
  \\ a)  The lowest order
($K^0$)  field equation is in flat space case  just  the ordinary massless
 Dirac
equation  for the  induced spinors and classical field equations for  the
 extremals of
K\"ahler action.   $\Lambda^k$ vanishes in this order. \\ b) Substituting
 the solutions
of $K^0$  equations to field equations and writing them in first order in
 $K$
one obtains  Dirac equation for $\Psi^1$ with source term coming from the
currents $J^k$ defined by  $\Psi^0$.  $\Psi_1$ can be expressed as a
convolution
of propagator with the source term just as in ordinary quantum field
theory.  \\
c) One can continue the iteration in  similar manner but in practice few
lowest
orders should give the essentially exact solution for physically
interesting
values of $p$ of order $10^{38}$.  The condition for convergence is that
gradients of $CP_2$ coordinates are of order $K$ at most.

 \subsection{  Perturbative construction for the solution of field equations
 for nonwarped $M^4$}

Flat space background  theory should be all what is needed for the
application
of the theory in particle physics context.   Although the  gravitational
 warping of
$M^4$ must be taken into account in realistic theory it is useful to look
the
construction theory first for the standard imbedding of $M^4$.

\vm

In the lowest order one obtains just the
  massless  Dirac equation  in the  background determined by the classical
 induced
 spinor connection of $CP_2$. In the  standard $M^4$ background one has

\begin{eqnarray}
 \gamma^k \partial_k\Psi &=&0
 \end{eqnarray}

 \noindent  The most natural basis  for the spinors is the  p-adic
 planewave basis. The use of planewaves is natural since p-adic convergence
cube is the analog of  the finite string world sheet obtained by the
logarithmic
mapping $z\rightarrow w=ln(z)$ from infinite complex plane and $M^4$
coordinates
are just the counterpart of complex world sheet coordinate
$w=\tau+i\sigma$.

\vm

The  next order in $p$ introduces interactions and the Dirac equation
 contains gauge potential term given   by the super part of the $CP_2$
 coordinates
defined by $\Psi_0$.  Also the induced metric contains  term, which however
gives only a total divergence to the action in accordance with the idea that
gravitional interaction is weak and therefore $O(p^2)$ effect.   One
obtains
formally  massless Dirac equation for $\Psi_1$  so that  $\Psi_0$ defines a
source term.  One can solve $\Psi_1$  as a convolution of  the free field
propagator
defined by quantized $\Psi_0$ with source term.  One must however take this
argument very cautiously.  The  normal ordering of the  oscillator
operators  in the  source term gives via anticommutators a term linear in
oscillator
operators and  that mass term might be  generated.

\vm

The procedure can be continued.  In fourth order in $p$ (order is simply
the  maximal number of fermion currents in interaction term)  K\"ahler
 action
density gives nontrivial contribution to  the interaction Lagrangian.  This
contribution can be interpreted also as   EYM action for the  induced
gauge field and
metric  as proposed in \cite{TGD}:  the  difference with
respect to the  standard theory is that quantum gauge fields are fermion
 antifermion
composites.

\vm

There are good reasons to expect that
 perturbation expansion leads to the replacement of  the planewaves
$f_P$  with  their quantized counterpart obtained by replacing $M^4$
coordinate
with its super symmetric version.  In fact, if one takes $M^4$ super
coordinates as coordinates instead of $M^4$ coordinates one obtains
automatically super symmetrized planewaves.  The exponential is just the
 TGD:eish
counterpart for the  standard vertex operators appearing in string model
and n-point
functions  for various fields automatically contain this  kind of vertex
operator.

 \subsection{The effect of gravitational warping}

Flat space approximation works for the vacuum expectation value of the
 bosonic
field equations only. The general matrix element of the  bosonic field
equation
implies  the dependence of $H$ coordinates on various conserved
quantum numbers. The most important of these quantum numbers is 4-momentum.
The previous considerations suggest that standard  imbedding of $M^4$  is
 simply
replaced with a warped version of $M^4$.

\vm

Gravitational warping  is indeed necessary to make the loop
summations appearing in perturbation theory well defined and to remove
propagator poles.  This is achieved  assuming that\\ i)   propagator is
defined
in  terms of warped metric \\ ii) the plane wave state basis
is defined as state basis associated with ordinary massless Dirac operator,
to
which Dirac operator reduces in lowest order approximation.\\
  The use of
standard planewave basis is in principle possible.  A possible  good reason
for using this state basis is that the planewave basis associated  associated
warped metric does not satisfy the required ortogonality relations unless one
scales the sides of p-adic convergence cube in appropriate manner.

\vm

For the most  general
gravitational warping Dirac operator contains  mass term proportional to
second
fundamental form $H^k$:

\begin{eqnarray}
p^{\alpha}\gamma_{\alpha} \rightarrow
m^{\alpha\beta}p_{\beta}+ \frac{H^k\Gamma_k}{2}
H^k&=& D^{\alpha}\partial_{\alpha}h^k
\end{eqnarray}

\noindent If  this term is covariant
constant mass term can be included in propaator.  Covariant constancy  means
that  the $CP_2$ projection of warped $M^4$   can be regarded
as a path of particle in covariantly constant force field in $CP_2$. A
 little
calculation shows that second fundamental form is proportional to $P^2$ for
gravitational warping and therefore vanishes for massless states.

\section{Some open questions}

It is good to list some of the open questions not explicitely mentioned
 in the text.

\subsection{Description of hadrons as p-adic string like objects?}

The basic assumption of p-adic approach is super conformal invariance
for boundary components.  Hadron  mass calculations of fourth paper
of \cite{padmasses}
 are
based on general features of Super conformal invariance and doesn't exclude
either the  description of hadron as 'bag' or as a  string. Therefore one
 must
consider also the possibility of more general spacetimes such as string like
objects suggested in basic TGD as description of hadrons \cite{TGD}.
The formulation of the p-adic  field
 theory is  formally  possible arbitrary spacetime topology and approximate
 symmetries
are same as in $M^4\rightarrow CP_2$ provided $CP_2$ contribution to metric
is
small.   A particularly interesting possibility is the description of
 hadrons as
string like objects.  In this case the basic vacuum  topology would be
 $X^2\times
Y^2\in  M^4\times CP_2$,  $Y^2$ being   Lagrange manifold and $X^2$
two-dimensional string like object in $M^4$.   These surfaces are vacuum
extremals and canonical transformations produce new vacua of this type.
More
general vacua are obtained as maps $X^2\times Y^2$ to $M^4$ and physical
 hadrons
should correspond to 'thickened' strings with transversal dimension of
 order
hadronic length.     Quarks and gluons would correspond to boundary
 components
for  this surface.  One  obtains string model with dynamically generated
 string
tension coming from the energy associated with the super part of
$H$-coordinates.  What is however important that the general features of
hadronic mass spectrum are determined to high degree from the p-adic
thermodynamcis of boundary components and universality for critical systems
suggests that the predictions of string model description do not differ much
from the predictions of 'bag' model description.

 \subsection{  Kac Moody spinor concept}

There are good reasons to hope that the treatment of  the boundary
 components by
idealizing them with point like objects and using second quantized Kac Moody
spinors  provides good practical description of elementary particle physics.
The justification of the  Kac Moody spinor concept from the properties of
p-adicized K\"ahler Dirac action remains however to be found.

\vm

 The main technical problem related to the second quantization of Kac Moody
spinors is the construction of the action of bosonic KM spinors on  the
fermionic
and bosonic KM spinors to make possible the concrete realization of minimal
 coupling
gauge action.

\subsection{The values  of length scale $L_0$ and K\"ahler
coupling strength}

In this work a consistency argument leading to correct values for the length
scale $L_0$ and K\"ahler coupling strength was represented but the actual
deduction of the value of  $L_0$  and K\"ahler coupling strength  from
Quantum TGD remains lacking.

\section{Appendix:  Conserved currents}

 The currents associated with conformal and canonical  transformation
can be
 deduced using standard formulas.  The general form of transformation is
given
by

\begin{eqnarray}
 h^k &\rightarrow& h^k+\delta h^k\nonumber\\
\Psi &\rightarrow & \Psi +\delta \Psi
\end{eqnarray}

\noindent The change of spinor field come from gauge rotation associated
 with the tranformation.   In $CP_2$ degrees of freedom there is also  a
term
proportional to the Hamiltonian of the canonical transformation acting as
$U(1)$
gauge trasformation. The details are given  in main text.

\vm

The change of super coordinate can be derived using the constraint equation
 $h^k_S= h^k +K\bar{\Psi}\Gamma^k(h^k_S)\Psi$.   A little calculation shows
that super part of coordinate suffers a mere spin rotation   so that one
 has

\begin{eqnarray}
\delta h^k_S&=& \delta h^k + K\bar{\Psi} Comm(\partial^l
\delta h^m_S\Sigma_{mn},
\Gamma^k) \Psi
\end{eqnarray}

\noindent From this consistency condition $\delta h^k_S$ can be solved by
writing $\delta h^k_S$ in series with respect to $K$

\begin{eqnarray}
\delta h^k_S&=& \sum K^n\delta h^k_{S,n} \nonumber\\
\delta h^k_{S,n} &=& \bar{\Psi} Comm(\partial^l\delta h^m_{S,n-1}
\Sigma_{mn},
\Gamma^k) \Psi\nonumber\\
\delta h^k_{S,0}&=& \delta h^k
\end{eqnarray}

\noindent  The bosonic contribution to  the current is given  by

\begin{eqnarray}
J^{\alpha}_K&=& J^{\alpha}_k\delta h^k_S\nonumber\\
J^{\alpha}_k&=&\frac{\partial  L_K}{\partial (\partial_{\alpha}
h^k_S(x))}\nonumber\\ &=& (T^{\alpha\beta}
 \partial_{\beta}h^lh_{kl}) - F^{\alpha\beta} J_{k l }
\partial _{\beta}h^l )
\delta h^k_S
\nonumber\\
T^{\alpha\beta}&=& \frac{1}{4g_K^2}(J^{\nu\alpha}J_{\nu}^{ \
\beta}-
g^{\alpha \beta} J^{\mu \nu} J_{\mu \nu}/4)\sqrt{g}\nonumber\\
\
\end{eqnarray}

\noindent Fermionic contribution to the current is given by

\begin{eqnarray}
J^{\alpha}_D&=& J^{\alpha}_1+J^{\alpha}_2\nonumber\\
J^{\alpha}_1&=& \frac{\partial L_D}{\partial (\partial_{\alpha} \Psi
(x))}\delta \Psi = \bar{\Psi}\Gamma^{\alpha} \delta \Psi \nonumber\\
J^{\alpha}_2&=& \frac{\partial L_D}{\partial
(\partial_{\alpha}h^k (x))}\delta
h^k_S \nonumber\\
&=& (T_D^{\alpha\beta }\partial_{\beta}h^k +
 \bar{\Psi}\Gamma_kD^{\alpha}\Psi ))\delta h^k_S \nonumber\\
T_D^{\alpha\beta}&=& \bar{\Psi}( \Gamma^{\alpha} D^{\beta}+
 \Gamma^{\beta}D^{\alpha})\Psi \sqrt{g}\nonumber\\
\delta \Psi&=& \partial_l j_k\Sigma^{kl} +k H
\end{eqnarray}

\noindent  In this formula $\delta \Psi$ contains only spin rotation term
and Hamiltonian term since the term proportional to spinor  connection drops
from the transformation formula.
 Similar formulas can be derived for boundary contributions.  The spin
 rotation term gives fermion conformal spin $1/2$.  Fermions possess also
canonical spin comming from spin rotation in $CP_2$ degrees of freedom and
canonical $u(1)$ charge, which is essentially fermion number
 (1/3 for quarks).

\end{document}